\documentclass[aps,prl,twocolumn,superscriptaddress]{revtex4-1}

\usepackage{graphicx}
\usepackage{bm}
\usepackage{mathtools}
\usepackage{xspace}

\usepackage{tabularx}
\usepackage{multirow}
\usepackage{xcolor}
\usepackage{bm}

\newcounter{firstbib}

\def\NiFe{LiNi$_{0.8}$Fe$_{0.2}$PO$_4$}
\def\NiFex{LiNi$_{1-x}$Fe$_x$PO$_4$}
\def\Ni{LiNiPO$_4$}
\def\Fe{LiFePO$_4$}

\begin{document}

\title{Tuning magnetoelectricity in a mixed-anisotropy antiferromagnet}
\date{\today}

\author{Ellen Fogh}
\affiliation{Laboratory for Quantum Magnetism, Institute of Physics, \'{E}cole Polytechnique F\'{e}d\'{e}rale de Lausanne (EPFL), CH-1015 Lausanne, Switzerland}
\affiliation{Department of Physics, Technical University of Denmark, DK-2800 Kongens Lyngby, Denmark}
\author{Bastian Klemke}
\affiliation{Helmholtz-Zentrum Berlin für Materialien und Energie, D-14109 Berlin, Germany}
\author{Manfred Reehuis}
\affiliation{Helmholtz-Zentrum Berlin für Materialien und Energie, D-14109 Berlin, Germany}
\author{Philippe Bourges}
\affiliation{Laboratoire Léon Brillouin, CEA-CNRS, CEA-Saclay, F-91191 Gif-sur-Yvette, France}
\author{Christof Niedermayer}
\affiliation{Laboratory for Neutron Scattering and Imaging, Paul Scherrer Institute, Villigen CH-5232, Switzerland}
\author{Sonja Holm-Dahlin}
\affiliation{Nano-Science Center, Niels Bohr Institute, University of Copenhagen, DK-2100 Copenhagen Ø, Denmark}
\author{Oksana Zaharko}
\affiliation{Laboratory for Neutron Scattering and Imaging, Paul Scherrer Institute, Villigen CH-5232, Switzerland}
\author{Jürg Schefer}
\affiliation{Laboratory for Neutron Scattering and Imaging, Paul Scherrer Institute, Villigen CH-5232, Switzerland}
\author{Andreas B. Kristensen}
\affiliation{Department of Physics, Technical University of Denmark, DK-2800 Kongens Lyngby, Denmark}
\author{Michael K. Sørensen}
\affiliation{Department of Physics, Technical University of Denmark, DK-2800 Kongens Lyngby, Denmark}
\author{Sebastian Paeckel}
\affiliation{Helmholtz-Zentrum Berlin für Materialien und Energie, D-14109 Berlin, Germany}
\author{Kasper S. Pedersen}
\affiliation{Department of Chemistry, Technical University of Denmark, DK-2800 Kongens Lyngby, Denmark}
\author{Rasmus E. Hansen}
\affiliation{Department of Photonics Engineering, Building 343, 2800 Kongens Lyngby,Denmark}
\author{Alexandre Pages}
\affiliation{Laboratory for Quantum Magnetism, Institute of Physics, \'{E}cole Polytechnique F\'{e}d\'{e}rale de Lausanne (EPFL), CH-1015 Lausanne, Switzerland}
\author{Kimmie K. Moerner}
\affiliation{Department of Physics, Technical University of Denmark, DK-2800 Kongens Lyngby, Denmark}
\author{Giulia Meucci}
\affiliation{Department of Physics, Technical University of Denmark, DK-2800 Kongens Lyngby, Denmark}
\author{Jian-Rui Soh}
\affiliation{Laboratory for Quantum Magnetism, Institute of Physics, \'{E}cole Polytechnique F\'{e}d\'{e}rale de Lausanne (EPFL), CH-1015 Lausanne, Switzerland}
\author{Alessandro Bombardi}
\affiliation{Diamond Light Source Ltd., Harwell Science and Innovation Campus, Didcot, Oxfordshire, OX11 0DE, UK}
\affiliation{Department of Physics, University of Oxford, Parks Road, Oxford, OX1 3PU, UK}
\author{David Vaknin}
\affiliation{Ames Laboratory and Department of Physics and Astronomy, Iowa State University, Ames, Iowa 50011}
\author{Henrik. M. Rønnow}
\affiliation{Laboratory for Quantum Magnetism, Institute of Physics, \'{E}cole Polytechnique F\'{e}d\'{e}rale de Lausanne (EPFL), CH-1015 Lausanne, Switzerland}
\author{Olav F. Sylju{\aa}sen}
\affiliation{Department of Physics, University of Oslo, P. O. Box 1048 Blindern, N-0316 Oslo, Norway}
\author{Niels B. Christensen}
\affiliation{Department of Physics, Technical University of Denmark, DK-2800 Kongens Lyngby, Denmark}
\author{Rasmus Toft-Petersen}
\affiliation{Department of Physics, Technical University of Denmark, DK-2800 Kongens Lyngby, Denmark}
\affiliation{European Spallation Source ERIC, P.O. Box 176, SE-221 00, Lund, Sweden}

\date{\today}
\maketitle

{\bf
Control of magnetization and electric polarization is attractive in relation to tailoring materials for data storage and devices such as sensors or antennae. In magnetoelectric materials, these degrees of freedom are closely coupled, allowing polarization to be controlled by a magnetic field, and magnetization by an electric field, but the magnitude of the effect remains a challenge in the case of single-phase magnetoelectrics for application. We demonstrate that the magnetoelectric properties of the mixed-anisotropy antiferromagnet \NiFex\ are profoundly affected by replacing a fraction of the Ni$^{2+}$ ions with Fe$^{2+}$ on the transition metal site. This introduces random site-dependent single-ion anisotropy energies and causes a lowering of the magnetic symmetry of the system. In turn, magnetoelectric couplings that are symmetry-forbidden in the parent compounds, \Ni\ and \Fe, are unlocked and the dominant coupling is enhanced by two orders of magnitude. Our results demonstrate the potential of mixed-anisotropy magnets for tuning magnetoelectric properties.}

Utilizing the magnetoelectric (ME) effect to electrically control magnetic states has far-reaching prospects in next-generation electronics \cite{spaldin2019,liang2021}. Currently realised applications are based on heterostructure composites \cite{pradhan2020}, combining layers with distinct bulk properties. Proposals for application of the ME effect in heterostructures are abound, including electric-field control of skyrmions \cite{Skyrmion}, magnetoelectric spin-orbit logic devices \cite{manipatruni2019,trier2022}, medical implants \cite{nan2017,gu2019,dong2020} and low-power-consumption ME random access memory \cite{hu2011,liu2013,matsukura2015}. As many of these proposals involve distinct ME layers, a fundamental understanding of the properties of single-phase magnetoelectrics is pivotal to their realization. While our understanding of the underlying mechanisms has been greatly improved since the discovery of the ME effect, the relatively weak ME couplings are lingering barriers for applicability of single-phase magnetoelectrics.

The ME properties of a given single-phase material are a consequence of the magnetic point group symmetry inherent to its magnetically ordered state \cite{eerenstein2006,rivera2009}. More specifically, the absolute and relative orientation of the ordered moments dictate the non-zero elements of the ME tensor describing the coupling between electric and magnetic degrees of freedom \cite{schmid2008,rivera2009}. We explore chemical tuning in mixed-anisotropy antiferromagnets as a novel route for tailoring the properties of single-phase magnetoelectrics. Mixing magnetic ions with incompatible, or mismatched, single-ion anisotropies gives rise to what can be thought of as a composite on the atomic level. This random site-dependent anisotropy in combination with the inter-species exchange interaction creates frustration in the system and may result in what is known as an \textit{oblique} antiferromagnetic phase. Here, the ordered moments are oriented away from any of the easy axes observed in the stoichiometric compounds \cite{fishman1978,oguchi1978,mano1990}.

\begin{figure*}
    \centering
    \includegraphics[width = 0.75\textwidth]{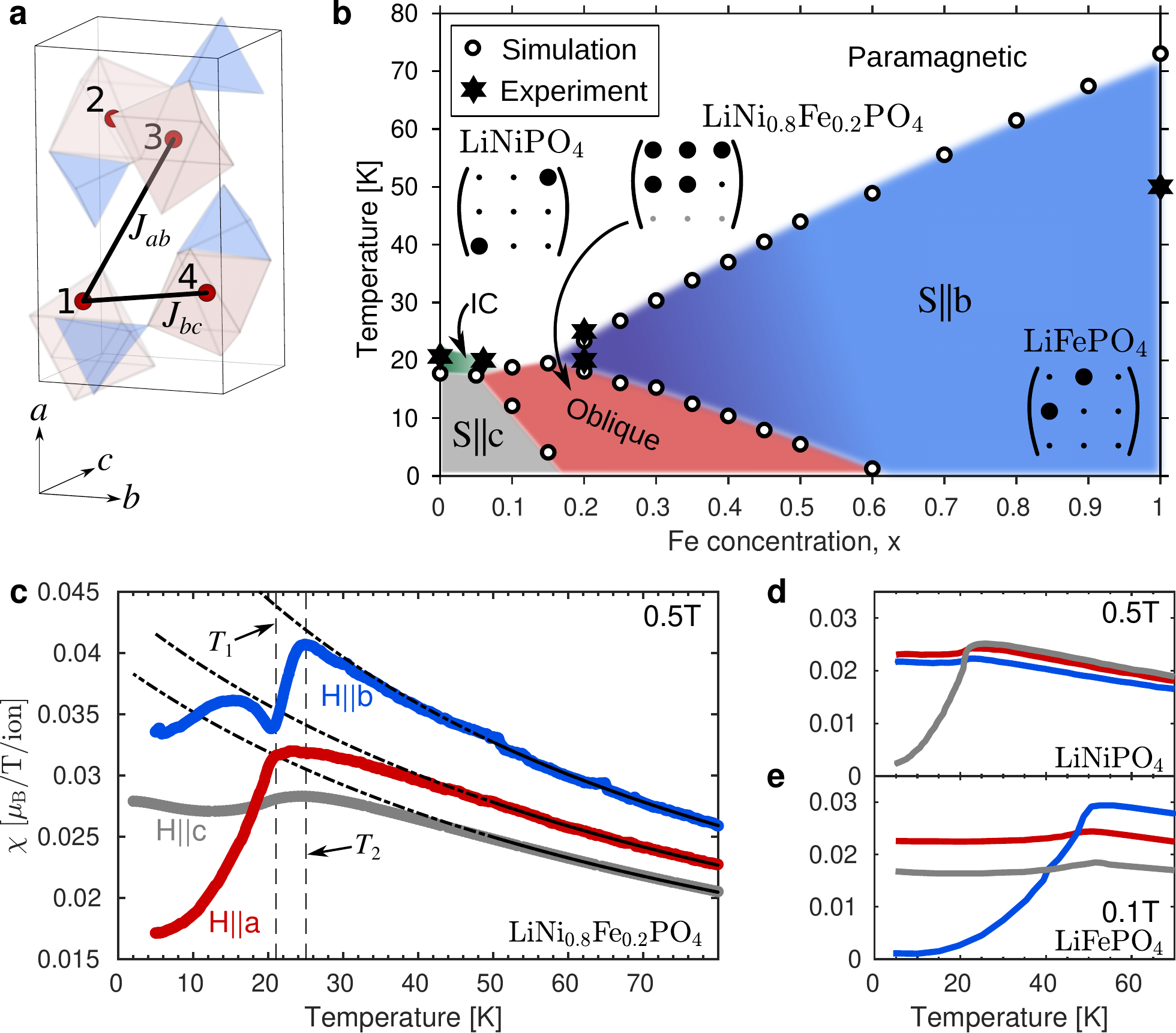}
    \caption{\textbf{Unit cell, $(x,T)$ phase diagram and magnetic susceptibility of \NiFex.} \textbf{a}, Crystallographic unit cell of Li$M$PO$_4$ with four magnetic ions (red numbered spheres) and the two most important exchange paths, $J_{bc}$ and $J_{ab}$, shown. The $M$O$_6$ octahedra and PO$_4$ tetrahedra are illustrated with red and blue shading, respectively.
    \textbf{b}, $(x,T)$ phase diagram constructed from experimental data and simulation. The open circles correspond to phase transitions observed in the simulated specific heat. Filled stars represent phase transitions detected in magnetic susceptibility and neutron diffraction experiments for samples with $x = 0,0.06,0.20$ and $1$. Both simulations and experiments reveal three phases: Commensurate phases with $S||c$ and $S||b$ are seen at small and large $x$, respectively, while an oblique phase is present in the range $0.1 < x < 0.6$. For each phase, the observed form of the magnetoelectric tensor at low temperature is indicated. The shading in the phase with $S||b$ illustrates that the ordered moment along $b$, $\langle S||b \rangle$, decreases when decreasing $x$ while $\langle S||a \rangle \approx \langle S||c \rangle = 0$. For small $x$ there exists an incommensurate (IC) phase in a narrow temperature interval above the commensurately ordered phase \cite{vaknin2004,jensen2009_struc}.
    \textbf{c}, Magnetic susceptibility of \NiFe\ measured with $0.5\,\mathrm{T}$ applied along the three crystallographic axes; $a$ (red curve), $b$ (blue curve) and $c$ (grey curve). Fits to the Curie-Weiss law are shown with black lines. The solid parts of the lines indicate the fitted interval ($50-300\,\mathrm{K}$) and the dash-dotted parts are extrapolations to lower temperatures. The vertical dashed lines mark transitions at $21$ and $25\,\mathrm{K}$. \textbf{d}-\textbf{e}, Corresponding susceptibilities for \Ni\ \cite{toftpetersen2011} and \Fe\ \cite{liang2008}, measured with $0.5$ and $0.1\,\mathrm{T}$ respectively.}
    \label{fig:phasediagram_susceptibility}
\end{figure*}

Our system of choice is \NiFex, which belongs to a well-known family of isostructural magnetoelectrics \cite{MercierThesis} with chemical formula Li$M$PO$_4$ ($M$ = Mn, Fe, Co, Ni) and space group \textit{Pnma} (No. 62). The crystallographic unit cell is illustrated in Fig. \ref{fig:phasediagram_susceptibility}a. The parent compounds, \Ni\ ($S = 1$) \cite{santoro1966,vaknin2004, jensen2009_struc} and \Fe\ ($S = 2$) \cite{santoro1967,li2006} order antiferromagnetically at $20.8\,\mathrm{K}$ and $50\,\mathrm{K}$, respectively. Below their N\'eel temperatures, they display similar commensurate spin structures except for the orientation of the magnetic moments, which are predominantly along the crystallographic $b$ and $c$ axes for respectively \Fe\ and \Ni. In \Ni\ there exists in addition an incommensurate phase in a narrow temperature interval just above the N\'eel temperature \cite{vaknin2004,jensen2009_struc}. The static and dynamic properties of Li$M$PO$_4$ are well-described by the spin Hamiltonian 
\begin{equation}
    \hat{\mathcal{H}} = \sum_{\langle i,j \rangle} J_{ij} {\bf S}_i \cdot {\bf S}_j + \sum_{i,\alpha} D^{\alpha}_i (S^{\alpha}_i)^2,
    \label{eq:Hamiltonian}
\end{equation}
where the first sum accounts for the exchange interactions of magnitude $J_{ij}$ between spins on sites $i$ and $j$. The second sum over all sites $i$ and three crystallographic directions, $\alpha = \left\lbrace a,b,c \right\rbrace$, reflects single-ion anisotropy energies, parameterized by the vector $D=(D^a,D^b,D^c)$. This term is responsible for the distinct ordered moment direction selected upon ordering in stoichiometric \Ni\ \cite{jensen2009_dyna} and \Fe\ \cite{toftpetersen2015}.

We have employed magnetic susceptibility measurements, neutron diffraction and Monte Carlo simulations to investigate the $(x,T)$ phase diagram of \NiFex\ (see Fig. \ref{fig:phasediagram_susceptibility}b). We observe three commensurate magnetic phases with propagation vector ${\bf k} = 0$. At low temperature and for $x < 0.2$, the spins order along $c$ like in \Ni. For $x > 0.6$, the spins order along $b$ like in \Fe. For $x = 0.2$, two magnetic phases appear upon cooling \cite{zimmermann2013}. Neutron diffraction reveals ordered moments predominantly along the crystallographic $b$-axis below $T_2 = 25\,\mathrm{K}$, while below $T_1 = 21\,\mathrm{K}$, the moments partially reorient towards the $a$-axis in a low-temperature oblique phase. Our investigations of the field-induced polarization in these phases have uncovered a complex ME coupling scheme. The lowered magnetic symmetry of the oblique phase combined with the broken discreet translational symmetry, unlocks ME tensor elements that are otherwise forbidden in the parent compounds.
Simulations show that the key factors responsible for the observed oblique phase are mismatched anisotropies combined with an inter-species exchange coupling creating competing exchange and single-ion anisotropy energy terms.
This unusual mechanism is of general applicability and represents a promising route to search for oblique ME phases in other families of compounds where the ME properties can be chemically tuned.

\bigskip
\noindent
{\bf \large Results}
\medskip

\noindent {\bf Magnetic susceptibility.} Figures \ref{fig:phasediagram_susceptibility}c-e illustrate distinct differences in magnetic susceptibility between \NiFe\ and its parent compounds, \Ni\ and \Fe. The susceptibility curves, $\chi_a$, $\chi_b$ and $\chi_c$, of both \Fe\ and \Ni\ for fields along $a$, $b$ and $c$ display textbook behavior for antiferromagnets with easy axes along $b$ and $c$, respectively. The component of $\chi$ parallel to the easy axis drops towards zero below the transition temperature while the two perpendicular components remain nearly constant. By contrast, the susceptibility of \NiFe\ shows clear evidence of two magnetic phase transitions. Below $T_2=25\,\mathrm{K}$, $\chi_b$ decreases while $\chi_a$ and $\chi_c$ remain constant. At a slightly lower temperature, $T_1 = 21\,\mathrm{K}$, $\chi_a$ begins to drop precipitously and the decrease of $\chi_b$ is interrupted, while $\chi_c$ remains approximately constant. These observations are indicative of a negligible $c$-axis component of the ordered moment at all temperatures, and of a rotation of the ordered moments from the $b$ axis towards the $a$ axis for temperatures lower than $T_1$. These two transitions were previously reported and we compare our findings with those of the authors of Ref. \cite{zimmermann2013} later in the Results section. Note that overall the susceptibility of the mixed system is higher than for the parent compounds. This, together with the overall different temperature dependence of the susceptibility as compared to the parent compounds, is evidence that \NiFe\ is indeed a solid solution and we can exclude phase separation in the system.

\begin{figure}
	\centering
	\includegraphics[width = \columnwidth]{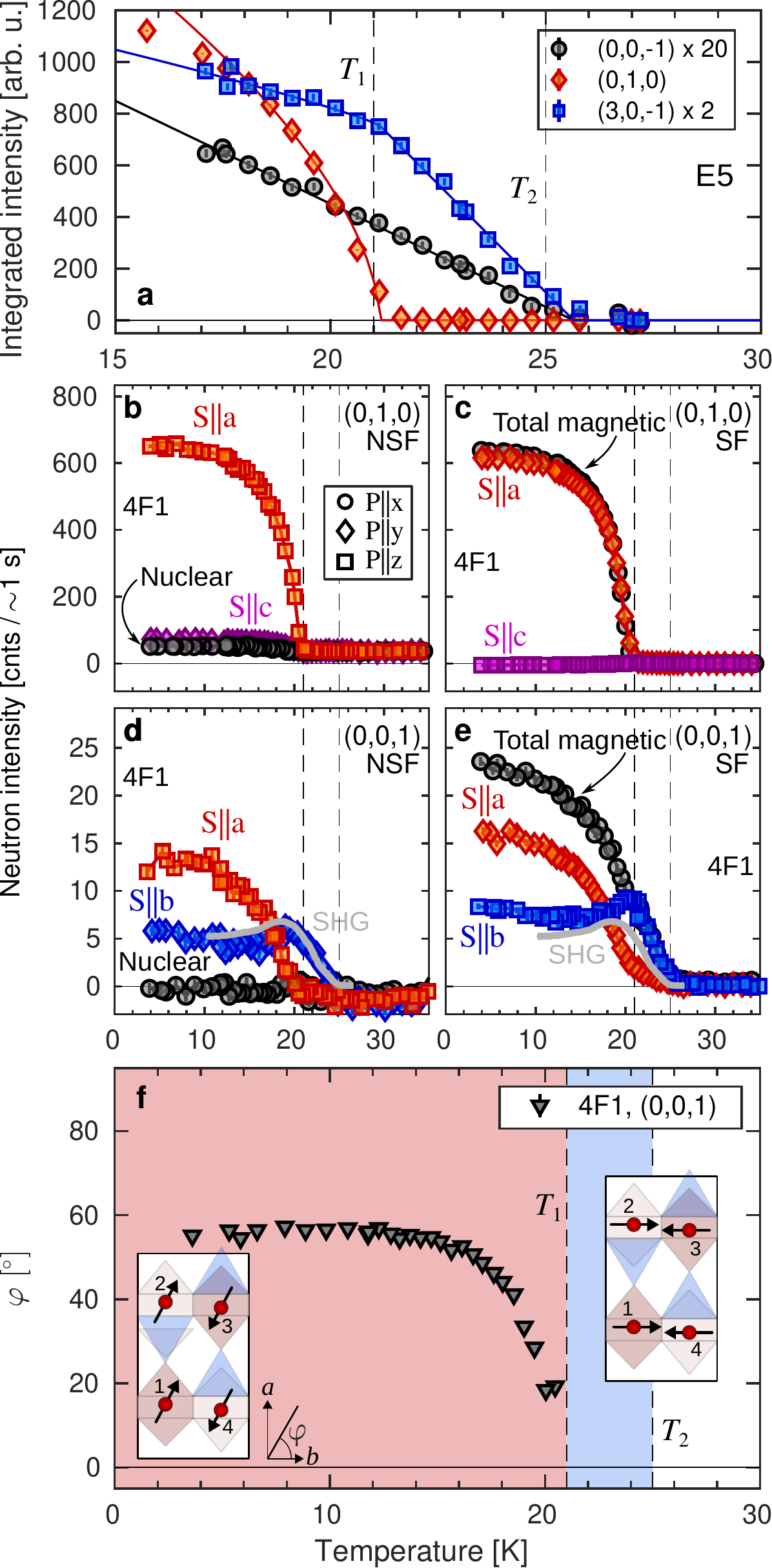}
	\caption{\textbf{Neutron diffraction.} \textbf{a}, Temperature profiles of neutron diffraction intensities for three magnetic Bragg peaks. Intensities are scaled as indicated in the legend. The solid lines show the fits described in the main text. \textbf{b}-\textbf{e}, Peak intensities obtained with polarized neutrons in the NSF (panels \textbf{b} and \textbf{d}) and SF (panels \textbf{c} and \textbf{e}) channels after correcting for non-perfect beam polarization \cite{stewart2009}. The crystal was oriented with the $a$ axis vertical such that NSF intensities for $P||z$ reflect $a$ axis spin components for both $(0,1,0)$ and $(0,0,1)$. Grey curves in panels \textbf{d}-\textbf{e} show optical second harmonic generation measurements from Ref. \cite{zimmermann2013}. \textbf{f}, Moment rotation angle, $\varphi$, as a function of temperature with projections in the $(0,0,1)$ plane of the magnetic structures for $T \leq T_1$ and $T_1 \leq T \leq T_2$. The two transitions at $T_1$ and $T_2$ are marked by vertical dashed lines in all panels.}
    \label{fig:neutron}
\end{figure}

\noindent {\bf Magnetic structures.} 
To determine the magnetic structures in \NiFe\ we turn to neutron diffraction. At all temperatures below $T_2$, the commensurate magnetic Bragg peaks were found to be resolution limited, implying long-range order (see Supplementary Information). 
A representative selection of temperature-dependent integrated intensities as obtained at the diffractometer, E5, is shown in Fig. \ref{fig:neutron}a.
The intensity of each magnetic Bragg peak reflects different combinations of symmetry components of the magnetic order. In addition, it carries information about the spin orientation in the ordered states, because neutrons couple exclusively to components of the magnetic moment perpendicular to the scattering vector $\bf Q$ (see Table \ref{tab:structurefactors} in the Supplementary Information). Our analysis indicates that the main magnetic structure component at all temperatures below $T_2$ is $(\uparrow \uparrow \downarrow \downarrow)$ with the numbering of spins defined in Fig. \ref{fig:phasediagram_susceptibility}a. Rietveld refinement of the magnetic Bragg peak intensities at base temperature yields magnetic moments predominantly in the $(a,b)$-plane with major component along $a$. For $T_1 \leq T \leq T_2$, our data suggests moments aligned along $b$.

The two transitions observed in our susceptibility measurements have clear signatures in the diffraction data: The $(0,0,-1)$ and $(3,0,-1)$ reflections grow linearly with decreasing temperature below $T_2 \sim 25\,\mathrm{K}$. By contrast, the $(0,1,0)$ peak appears only below $T_1 \sim 21\,\mathrm{K}$ where in addition, there is a kink in the temperature profile of the $(3,0,-1)$ intensity. The temperature dependencies of all recorded peaks are well described by a combination of a linear function and a power law, reflecting the existence of two order parameters, below $T_2$ and $T_1$, respectively (solid lines in Fig. \ref{fig:neutron}a and Fig. \ref{fig:E5complete} in the Supplementary Information). Simultaneous fits to all data sets yield transition temperatures $T_2 = 25.7(2)\,\mathrm{K}$ and $T_1 = 20.8(1)\,\mathrm{K}$ respectively, in good agreement with Refs. \cite{zimmermann2013} and \cite{li2009}. We note that the critical exponents for the two order parameters are clearly different. Below $T_2$, the neutron intensity increases linearly with decreasing temperature which means a critical exponent of $\frac{1}{2}$ as assumed fixed in the fit. This corresponds to the critical exponent resulting from long-range interactions or from a secondary order parameter. At $T_1$, $(0,1,0)$ displays a power law behavior with $\beta = 0.32(3)$ which is comparable to the critical exponent of a 3D Heisenberg, XY or Ising system.


To unambiguously determine the spin orientations, we performed a polarized neutron diffraction experiment using the triple axis spectrometer 4F1 and with scattering vector ${\bf Q} = (0,K,L)$ in the horizontal scattering plane. Uniaxial polarization analysis allows the two spin components perpendicular to $\bf Q$ to be individually addressed. This is done by measuring spin-flip (SF) and non spin-flip (NSF) intensities for the neutron beam polarization along the scattering vector ($P||x$), perpendicular to $\bf Q$ in the horizontal scattering plane ($P||y$), and along the direction perpendicular to the scattering plane ($P||z$). The temperature-dependencies of the resulting six cross sections were collected for the $(0,1,0)$, $(0,0,1)$ and $(0,1,2)$ reflections. The SF cross sections carry information on spin components perpendicular to both $\bf Q$ and the neutron beam polarization $\bf P$. The NSF cross sections reveal spin components perpendicular to $\bf Q$ but parallel to $\bf P$ in addition to any finite nuclear Bragg peak intensity.

Noting that the $(0,1,0)$ magnetic peak exclusively reflects $(\uparrow \uparrow \downarrow \downarrow)$ symmetry components (see Supplementary Information), Figs. \ref{fig:neutron}b-c show that the magnetic structure below $T_1$ involves sizeable spin components along $a$, but only negligible $c$-axis components. Spin components parallel to $b$ do not contribute to magnetic scattering at ${\bf Q} = (0,1,0)$, but can be probed at ${\bf Q} = (0,0,1)$ or $(0,1,2)$. Figs. \ref{fig:neutron}d-e confirm the involvement of an $a$-axis spin components below $T_1$, and show that the scattering is dominated by spins oriented along $b$ in the range $T_1 \leq T \leq T_2$. Note that here we plot only data for $(0,1,0)$ and $(0,0,1)$ as their interpretation is straightforward. The data for $(0,1,2)$ is shown in the Supplementary Information. A comparison of the observed intensities to the structure factors for the magnetic symmetry components contributing to the $(0,1,0)$, $(0,0,1)$ and $(0,1,2)$ peaks makes it clear that the dominant symmetry component for $T_1 \leq T \leq T_2$ is also $(\uparrow \uparrow \downarrow \downarrow)$. The scattering from $b$-axis spin components, reflected by the NSF, $P||y$ and SF, $P||z$ cross sections in Figs. \ref{fig:neutron}d-e increases monotonically for temperatures in the range $T_1 \leq T \leq T_2$ and levels off to a finite value at our experimental base temperature. The rotation angle, $\varphi$, in the $(a,b)$-plane may be calculated from the ratio of $P||y$ and $P||z$ data in Figs. \ref{fig:neutron}d-e leading to the conclusion that the angle between the moments and the $b$ axis approaches $\varphi=60^{\circ}$ at low temperatures, see Fig. \ref{fig:neutron}f. 

The small but finite nuclear intensity for $P||x$ in Fig. \ref{fig:neutron}b and Fig. \ref{fig:4F1scans}a may be due to a change of the lattice symmetry which could be caused by magnetostriction. Magnetostriction is common in magnetoelectrics and for \Fe\ this effect has been observed when applying magnetic fields \cite{werner2019}. Future synchrotron X-ray studies will uncover the evolution of the crystal lattice and symmetry as a function of temperature.

\begin{figure*}
	\centering
	\includegraphics[width = \textwidth]{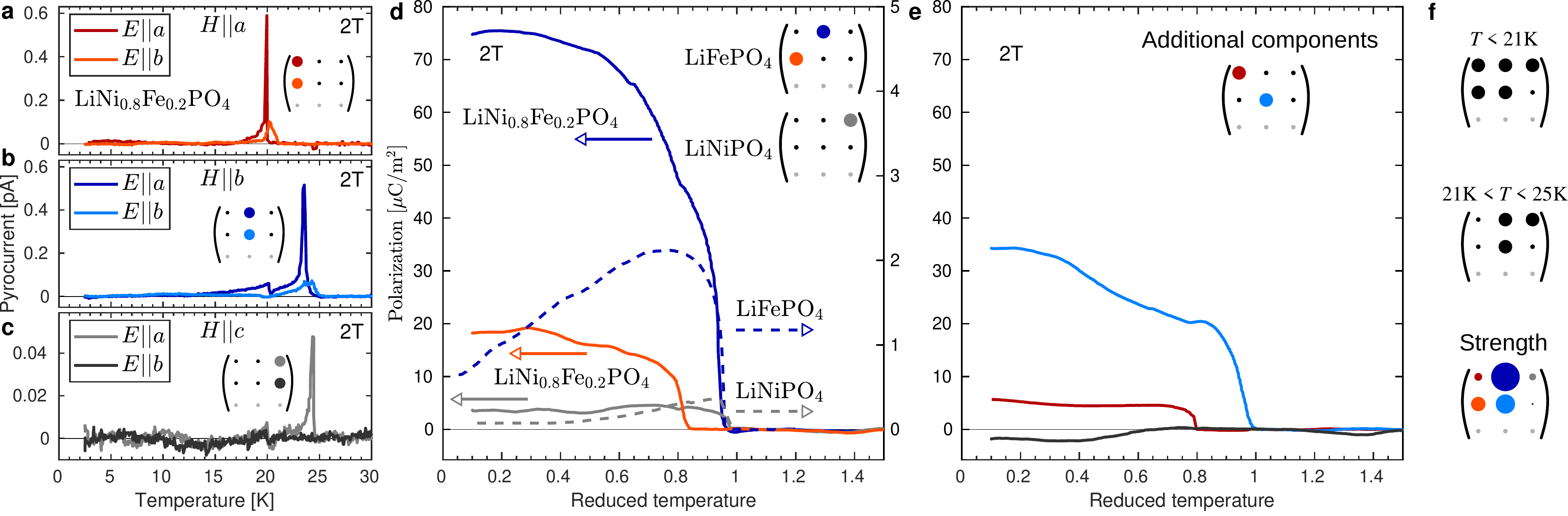}
	\caption{\textbf{Pyrocurrent and magnetoelectric effect.} Panels \textbf{a}-\textbf{c} show the pyrocurrent for \NiFe\ as a function of temperature for magnetic fields applied along $a$, $b$ and $c$, respectively. The insets indicate which elements of the ME tensor, $\alpha$, were probed. \textbf{d}, Electric polarization as a function the reduced temperature with transition temperatures $21$, $25$ and $50\,\mathrm{K}$ at zero field for \Ni, \NiFe\ and \Fe, respectively. Note the two $y$-axes: the left one for the data for the mixed system and the right one for the parent compounds. \textbf{e}, Temperature dependency of the electric polarization originating from tensor elements not present in the parent compounds of \NiFe. All measurements shown here were carried out with an applied magnetic field strength of $2\,\mathrm{T}$ where the ME effect is still linear. Panel \textbf{f} gives an overview of the observed non-zero ME tensor elements for $T < T_1$ and $T_1 \leq T \leq T_2$ as well as the relative strength of the elements in the oblique phase.}
	\label{fig:polarization}
\end{figure*}

The solid grey lines in Figs. \ref{fig:neutron}d-e represent the intensity of the second harmonic generation (SHG) susceptibility tensor element, $\chi_{zxx}$, from Ref. \cite{zimmermann2013}. Here the first subscript signifies the component of the non-linear polarization induced by an electric field with components denoted by the last two subscripts. The similarity of the SHG signal with the NSF, $P||y$ and SF, $P||z$ cross sections is clear evidence that these two observations are intimately related. The SHG data was interpreted by the authors of Ref. \citep{zimmermann2013} as a signature of spin rotation from the easy $b$ axis of stoichiometric \Fe\ towards the easy $c$ axis of stoichiometric \Ni, upon cooling below $T_1$. Our polarized neutron diffraction results only allow for a small spin component along $c$ and show instead a sizeable component along $a$. This picture is consistent with the susceptibility data in Fig. \ref{fig:phasediagram_susceptibility}c. The physical mechanism for this surprising reorientation away from the easy axes of the two parent compounds is explored in our Monte Carlo simulations to be presented further on, but first we explore its profound consequences for the ME coupling.

\noindent {\bf Magnetoelectric effect.} 
The linear ME effect is described by the relation ${\bf P}^{\mathrm{E}} = \alpha_{ij} {\bf H}$ between the components of the induced electrical polarization, ${\bf P}^{\mathrm{E}}$, and those of the applied magnetic field, $\bf H$. A related equation, $\mu_0 {\bf M} = \alpha^T {\bf E}$, connects the components of the induced magnetization, $\bf M$, to those of the applied electric field, $\bf E$. For systems invariant to integer lattice vector translations, the allowed elements of the ME tensor $\alpha$ are imposed by the point group symmetry of the magnetically ordered state \cite{rivera2009,schmid2008}. Specifically, for the stoichiometric parent compounds \Ni\ and \Fe, the reported magnetic structures imply that the elements which may be non-zero are $\alpha_{ac}$, $\alpha_{ca}$ and  $\alpha_{ab}$, $\alpha_{ba}$, respectively.

The ME response of \NiFe\ was probed with measurements of the pyrocurrent produced by a temperature change (see Methods and the Supplementary Information for details). Our results for \NiFe\ are shown in Fig. \ref{fig:polarization} and are compared to the ME response of the parent compounds, \Ni\ and \Fe\ \cite{fogh2022_Fe}. Note that in the following analysis we assume space group \textit{Pnma}. However, it was recently shown that LiFePO$_4$ may display a lower symmetry \cite{fogh2022_Fe}. The pyrocurrent for \NiFe\ for two orthogonal orientations of the electric poling field, $\bf E$, and three directions of the magnetic field shows clear signatures of two ME phase transitions slightly below $T_2$ and $T_1$, see Figs. \ref{fig:polarization}a-c. The evidence is in the form of spikes in the pyrocurrent, which following a geometrical correction can be integrated to obtain the temperature dependent polarization components $P^{\mathrm{E}}_i$. 

The electric polarization corresponding to the tensor elements $\alpha_{ab}$ and $\alpha_{ba}$ together with that corresponding to $\alpha_{ac}$ are shown in Fig. \ref{fig:polarization}d. As mentioned above, these components are known to be non-zero for stoichiometric \Fe\ and \Ni\ \cite{santoro1967,mercier1968}. When comparing the ME response of \NiFe\ to that of \Fe\ and \Ni\ measured under identical conditions (blue and grey dashed lines in Fig. \ref{fig:polarization}d), it is apparent that the polarizations induced along $a$ and $b$ are significantly larger in \NiFe\ at all temperatures below the transition temperature. Most strikingly, in the limit $T \rightarrow 0$, the polarization due to the dominant tensor component, $\alpha_{ab}$ is increased by two orders of magnitude compared to \Fe. A second remarkable observation is that the onset temperatures of $\alpha_{ab}$ and $\alpha_{ba}$ are different. $\alpha_{ba}$ vanishes in the range $T_1 \leq T \leq T_2$ whereas $\alpha_{ab}$ is finite already below $T_2$ and displays a kink at $T_1$.

Finally, in Fig. \ref{fig:polarization}e we probe tensor components that are by symmetry not allowed for \Ni\ and very small for \Fe \cite{fogh2022_Fe}. Similarly, we observe $\alpha_{bb}$ below $T_2$ while $\alpha_{aa}$ is finite only below $T_1$. For the last tensor element measured, $\alpha_{bc}$, there is no spike to be seen in the pyrocurrent and we conclude that this element is either very weak or zero.

\begin{figure}
	\centering
	\includegraphics[width = \columnwidth]{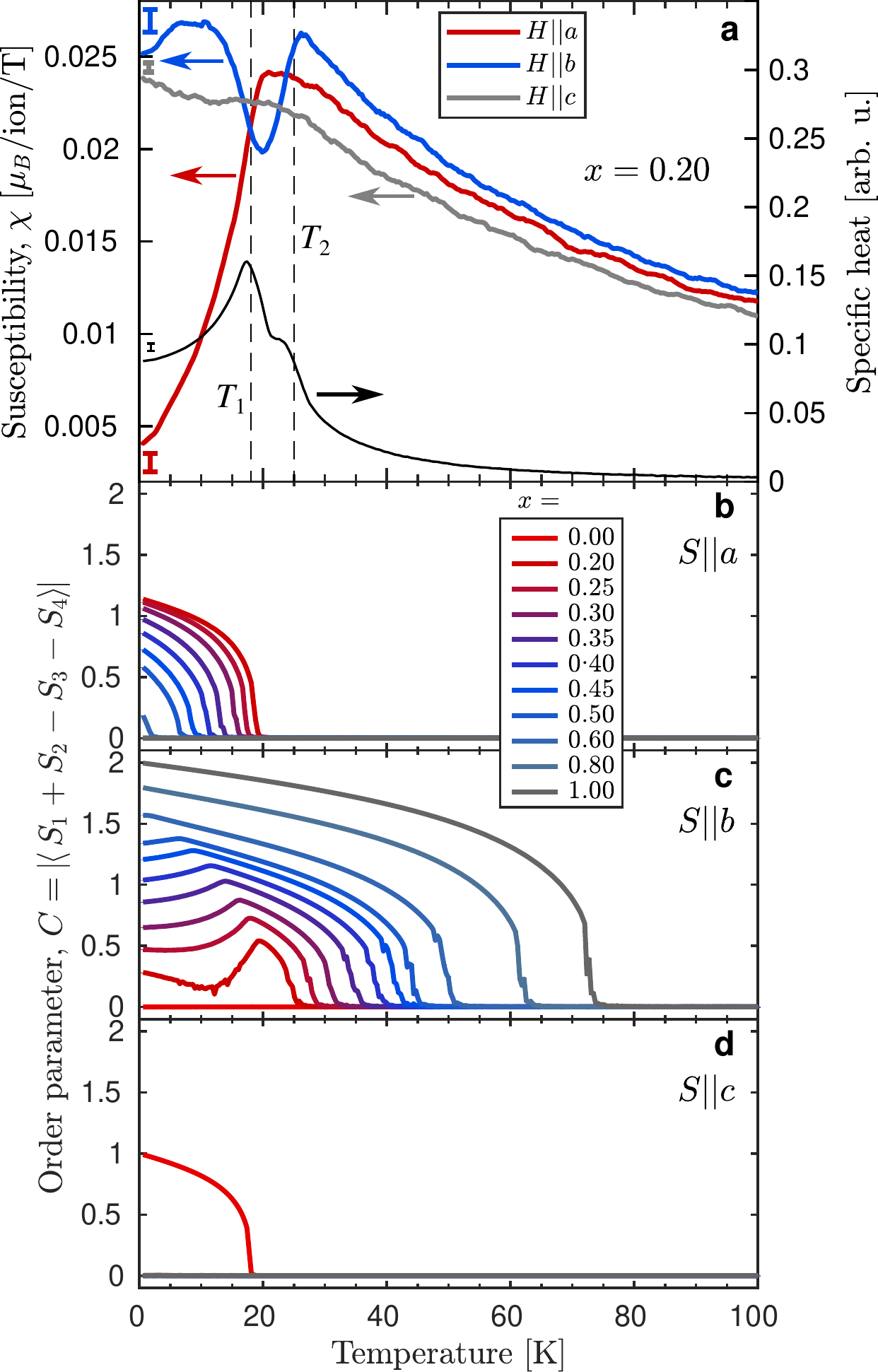}
	\caption{\textbf{Monte Carlo simulations.} \textbf{a}, Magnetic susceptibility (left axis) and specific heat (right axis) for $x = 0.20$. The maximum error bar sizes are indicated at the lowest temperatures. \textbf{b}-\textbf{d}, The absolute value of the $C$-type order parameter plotted as a function of temperature for spin components along the three crystallographic axes and for different values of $x$. Note that in \textbf{d} the curves for $x \geq 0.20$ coincide.}
	\label{fig:simulation}
\end{figure}

\noindent {\bf Monte Carlo simulations.} 
We now show that classical Monte Carlo simulations reproduce the salient features of the susceptibility and diffraction results for $x = 0.20$. In the calculations, we chose $J_{bc} = 1$ meV and $J_{ab} = 0.3$ meV for all corresponding pairs of sites (see Fig. \ref{fig:phasediagram_susceptibility}a), irrespective of their occupancy by Ni or Fe. For the single-ion anisotropies we used $(D^a,D^b,D^c) = (0.3,1.8,0)\,\mathrm{meV}$ for Ni with $S = 1$ and $(D^a,D^b,D^c) = (0.6,0,1.6)\,\mathrm{meV}$ for Fe with $S = 2$. This choice of parameters looks like a dramatic simplification when compared to the full set of experimentally determined parameters \cite{jensen2009_dyna, toftpetersen2011,toftpetersen2015} for \Ni\ and \Fe\ as given in Table \ref{tab:parameters} of the Supplementary Information. Nevertheless, the set of simplified parameters yields results in good agreement with our experimental neutron diffraction and susceptibility measurements for $x=0.20$. In turn, this agreement justifies the use of the Hamiltonian from Eq. \eqref{eq:Hamiltonian} with the chosen parameters for an exploration of the full phase diagram of \NiFex.

In Fig. \ref{fig:simulation}a, we plot the calculated magnetic susceptibility and specific heat for $x=0.20$ as a function of temperature. Two phase transitions are observed near $25$ and $20\,\mathrm{K}$. The transition temperatures as well as the temperature-dependencies of the three components of $\chi$ are in excellent agreement with the experimental results shown in Fig. \ref{fig:phasediagram_susceptibility}c. The accuracy of the simulations is further illustrated by comparing the calculated and measured susceptibilities for $x = 0.06$ with the corresponding experimental data (see Supplementary Information).

Next, we study the simulated $C$-type order parameter $C = \left| \left\langle S_1 + S_2 - S_3 - S_4 \right\rangle \right|$ for spin components along $a$, $b$ and $c$, respectively. Figures \ref{fig:simulation}b-d show the temperature dependencies of $C$ for the full range of Fe concentrations, $x$. Focusing on $x = 0.20$, the resemblance with the polarized neutron diffraction data in Fig. \ref{fig:neutron} is striking. Note that the neutron diffraction intensity is proportional to the order parameter squared. The first phase transition at $25\,\mathrm{K}$ corresponds to spins ordering along $b$. The growth of the corresponding order parameters is interrupted at $20\,\mathrm{K}$ where a rotation towards $a$ starts and the oblique low-temperature phase is entered. The $c$-axis component remains zero at all temperatures. From the $a$ and $b$ components of the simulated order parameter we arrive at a rotation angle of $76^{\circ}$ at low temperature, which compares reasonably well with the value of $\sim60^{\circ}$ obtained from the experimental data, re-visit Fig. \ref{fig:neutron}f. We use the transition temperatures derived from the simulated specific heat and order parameters data to construct the $(x,T)$ phase diagram shown in Fig. \ref{fig:phasediagram_susceptibility}b. The simulations underestimate the transition temperatures for small $x$ compared to the measured values (star symbols for $x = 0$, $0.06$ and $0.20$), but the ratios of simulated to observed transition temperatures are relatively constant with $x$ in this range.

The oblique antiferromagnetic state is relatively robust. The simulations show that the only requirements are an inter-species exchange interaction as well as competing single-ion anisotropies with opposite easy and hard axes for the parent compounds but a common intermediate axis off any of these easy and hard axes. It is this frustration between exchange and single-ion anisotropy energies that generates the oblique state. In the analysis of the neutron diffraction data we assumed a collective behavior of all spins regardless of species. The simulations show that indeed the ensemble average of the moments give a collectively ordered picture. However, we also find local fluctuations between Ni and Fe sites (see Fig. \ref{fig:MCstructures} in the Supplementary Information). The ordered moment for the oblique phase is therefore lower when calculating the average over the entire system than when considering individual sites, not only due to thermal fluctuations but also due to site specific differences in the moment orientation. This issue and the general consequences of violation of discreet translational symmetry for the ME effect are an interesting topic of future theoretical investigations.

\bigskip
\noindent
{\bf \large Discussion and conclusions}
\medskip

\noindent 
The effects on the magnetic ground state of a quenched random distribution of ions in mixed anisotropy antiferromagnets have been extensively studied by renormalization group theory and in various mean field models \cite{aharony1975,fishman1978,lindgaard1976,matsubara1977,oguchi1978,mano1985,mano1990}. 
The resulting phase diagrams generically contain one or more oblique phases at intermediate compositions, in which the ordered moments are oriented away from the easy axes of the parent compounds. Depending on the details of the exchange and anisotropy terms in the spin Hamiltonian, the oblique ground state may involve ordered moments in a plane spanned by the easy axes of the two parent compounds, or perpendicular to both in the particular case where the easy and hard axis of one species coincide with the hard and easy axes of the other species \cite{oguchi1978}. These predictions of a chemically tunable magnetic ground state are broadly consistent with our experimental observations and Monte Carlo simulations for \NiFex{}, and have in the past been found to agree well with experimental studies of mixed-anisotropy antiferromagnets \cite{wong1980,katsumata1979,tawaraya1980,perez2015,bevaart1978,wegner1973}.

In the lithium orthophosphates, the magnetic $C$-type structure is the dominant structure component for all stoichiometric family members. The single-ion anisotropy plays a crucial role for the ME coupling, as the allowed tensor elements derive from the magnetic point group, \textit{ipso facto} the direction of the ordered moments. Since $S||b$ in LiFePO$_4$, $\alpha_{ab}$ and $\alpha_{ba}$ are allowed, while as $S||c$ in LiNiPO$_4$, $\alpha_{ac}$ and $\alpha_{ca}$ are allowed. When $S||a$, the diagonal elements are allowed ($\alpha_{aa},\alpha_{bb},\alpha_{cc} \neq 0$). It  follows from our results, however, that such simple rules do not apply in the mixed systems, where discreet translational symmetry is broken and the local spin anisotropy is site dependent. Between $21$ and $25\,\mathrm{K}$, the predominant spin orientation in \NiFe\ is $S||b$. Nevertheless, the observed diagonal tensor element $\alpha_{bb}$ is almost as strong as the expected $\alpha_{ab}$-component, while $\alpha_{ba} = 0$. Below $21\,\mathrm{K}$, the ME-tensor is more complex and most tensor elements are observed. This is due to the off-axis direction of the ordered moments. However, the fact that some expected tensor elements are extinct between $21$ and $26\,\mathrm{K}$, while some unexpected elements are not, is a strong hint that while discreet translational symmetry is broken, the existence of the ME coupling is still governed by magnetic point group symmetry, yet not in the same manner as in the stochiometric systems. This is an interesting point in itself and should be subject to further theoretical study.

The most intriguing observation is the hundredfold increase in the strength of the ME coupling for \NiFe\ observed in the pyrocurrent measurements. In LiNiPO$_4$ and LiFePO$_4$, the effect is believed to arise from exchange striction \cite{jensen2009_struc, toftpetersen2015,toftpetersen2017,fogh2020}. Specifically, the model connects the electric polarization caused by a displacement, $x_i$, of PO$_4$ tetrahedra along $i$, to the component of the applied magnetic field along $j$ as follows:
\[
    P^{\mathrm{E}}_i \propto x_i = \frac{\lambda_i}{\epsilon_i} \langle S \rangle^2 \chi_j H_j \, .
\]
Here $\langle S \rangle$ is the order parameter, $\chi_j$ the magnetic susceptibility for fields along $j$, $\epsilon_i$ the elastic energy constant associated with tetrahedron displacement ($\mathcal{E}_i = \epsilon_i x_i^2$) and $\lambda_i$ reflects the strength of the exchange striction ($J_{H \neq 0} \rightarrow J_{H = 0} + \lambda_ix_i$). In addition to the general increase in magnetic suceptibility in \NiFe\ as compared to the parents, both a reduction of the elastic displacement energy and an increase in exchange striction could cause stronger ME couplings.
A lowering of crystal symmetry may indeed result in lower energy cost for displacing the PO$_4$ tetrahedra as well as more allowed options for displacement directions. Moreover, the local number of nearest neighbors of a given species, and variations thereof, could bring exchange striction terms into play which would otherwise cancel out (in the parent compounds only interactions between ion pairs (1,2) and (3,4) contribute to the ME effect \cite{jensen2009_struc}). 


Tuning of the magnetic symmetry was also recently achieved in the olivine series of compounds, Li$_{1-x}$Fe$_x$Mn$_{1-x}$PO$_4$ \cite{diethrich2022}, as well as in Mn and Co doped LiNiPO$_4$ \cite{semkin2022}. These studies do not report on the corresponding consequences for the ME effect but they further illustrate that the lithium orthophosphate family harbour many possibilities for tailoring the magnetic and consequently also ME material properties. In general, the existence of ME and multiferroic oblique antiferromagnets is unlikely to be limited to this family. Notably, our Monte Carlo simulations show that competing single-ion anisotropies with a common intermediate axis combined with a significant inter-species exchange coupling are the decisive ingredients to prodcue an oblique magnetoelectric state, whereas details of the exchange coupling scheme play almost no role. Generally, transition metal ions exhibit complex single-ion anisotropies in octahedral environments, and will likely share an intermediate anisotropy axis in many families of compounds. An oblique ME phase may therefore also exist in other classes of materials. In future studies, Monte Carlo simulations similar to those performed in this work, possibly combined with DFT calculations to determine exchange constants, could precede time-demanding material synthesis in order to predict the viability of the candidate family to produce oblique ME phases. An interesting family of compounds for future studies of this type is for example Mn$_{1-x}M_x$WO$_4$ ($M = $Fe, Co, Cu, Zn) \cite{ye2008, ye2012, patureau2015, chaudhury2009}. 


In summary, we studied \NiFe\ experimentally using magnetometry, neutron diffraction and pyrocurrent measurements and theoretically through Monte Carlo simulations. We have identified an oblique low-temperature phase over an extended range of compositions. In this phase the spins rotate away from the distinct easy axes of the parent compounds, \Ni\ and \Fe, towards the direction of the common intermediate axis. The magnetoelectric properties correlate with the observed magnetic phase transitions, but the form of the magnetoelectric tensor departs from theoretical expectations based on discreet translational invariance and magnetic point group symmetries. Most dramatically, we observed a strong enhancement of two orders of magnitude for the dominant magnetoelectric tensor element compared to the parent compounds. These data in combination with our Monte Carlo simulation results suggest that the observations have broader implications and that chemical tuning of oblique magnetoelectric phases represents a promising path for tailoring magnetoelectric material properties.

\bigskip
\noindent
{\bf \large Methods}
\medskip

\noindent {\bf Magnetic susceptibility.} 
Magnetization measurements were performed using a Cryogenic Ltd. cryogen free vibrating sample magnetometer. A magnetic field of $0.5\,\mathrm{T}$ was applied along $a$ and $b$ for temperatures in the range $2-300\,\mathrm{K}$. 
For measurements with field along $c$ we used a Physical Property Measurement System (PPMS) from Quantum Design. The sample for all measurements of \NiFe\ was a $25\,\mathrm{mg}$ box-shaped single crystal. 

\noindent {\bf Neutron diffraction.} 
Unpolarized neutron diffraction data on \NiFe\ were collected at the E5 diffractometer at the Helmholtz-Zentrum Berlin using an Eulerian cradle, a neutron wavelength of $2.38\,\mathrm{Å}$ and a 2D position sensitive neutron detector. A pyrolytic graphite filter before the sample position was used for suppression of second order neutrons from the monochromator.

The polarized neutron diffraction data were obtained on the 4F1 triple-axis spectrometer at the Laboratoire Léon Brillouin, using a wavelength of $2.44\,\mathrm{Å}$. The monochromatic incident beam was polarized using a bending mirror and a pyrolytic graphite filter after the bender reduced second order contamination of the incident beam. A spin flipper was placed before the sample position. 
In combination with a Heusler analyzer this allowed for probing both spin flip and non spin-flip scattering. A set of Helmholtz-like coils around the sample position enabled polarization of the neutron beam along $x$, $y$ or $z$ in the coordinate system decribed in the Supplementary Information. 
The same high-quality $250\,\mathrm{mg}$ \NiFe\ single crystal was used in both polarized and unpolarized neutron diffraction experiments. At 4F1, it was oriented with $(0,K,L)$ in the horizontal scattering plane. Flipping ratios $F=40$ and $25$ were deduced from measurements on the structural $(020)$ and $(002)$ reflections, and used to correct for non-ideal beam polarization.

Preliminary studies of the magnetic structures of \NiFex\ were carried out at the triple-axis spectrometer RITA-II and TriCS at the Paul Scherrer Institute.

\noindent {\bf Pyrocurrent measurements.} 
The quasi-static method \cite{chynoweth1956} was used to perform pyrocurrent measurements with a Quantum Design PPMS at the Helmholtz-Zentrum Berlin. The custom-build insert for the measurement is described elsewhere \cite{paeckelBachelor}. Plate-shaped \NiFe\ single crystals with gold sputtered faces perpendicular to $a$ and $b$, and sample thickness $0.5$ and $0.9\,\mathrm{mm}$, respectively, were used. Single crystals of \Ni\ and \Fe\ were similarly prepared with faces perpendicular to $a$ and thickness $0.6$ and $0.5\,\mathrm{mm}$, respectively. A potential of $100\,\mathrm{V}$ was applied as well as a magnetic field while cooling to obtain a single ferroelectric domain. The potential was switched off at the experimental base temperature. The measurement was then performed upon heating at a constant temperature rate. Magnetic fields were applied along the $a$, $b$ and $c$ directions.

\noindent {\bf Monte Carlo simulations.} Classical Monte Carlo simulations were carried out using the spin Hamiltonian [Eq. \eqref{eq:Hamiltonian}] and employing the Metropolis algorithm \cite{metropolis1953}. The system was of size $10 \times 10 \times 10$ crystallographic unit cells (i.e. 4000 magnetic sites) and with Fe and Ni ions randomly distributed. The simulation was run for $10^5$ Monte Carlo steps for each temperature in the range $1-100\,\mathrm{K}$ with step size $0.35\,\mathrm{K}$ and decreasing temperature. 
For each temperature we use the final configuration from the previous temperature step as a starting point. This procedure mimics the process of slowly lowering the temperature in the physical experiment.
All simulations were run at zero field.

For each value of the Fe concentration, $x$, we simulated 5 distinct configurations from which we calculated the magnetic susceptibility, specific heat and order parameter. 
The susceptibility is calculated as $\chi = \beta \left( \left\langle M^2 \right\rangle - \left\langle M \right\rangle^2 \right)$, where $M$ denotes the total magnetization of the system, $\beta = \frac{1}{k_B T}$ and $k_B$ is the Boltzmann constant. The brackets, $\langle \rangle$, denote the ensemble average over system configurations. In Fig. \ref{fig:simulation}a and Fig. \ref{fig:chiLowFe}b we show $\chi$ normalized per magnetic site.
The specific heat is calculated from the energy dissipation theorem, $C_V = k_B \beta^2 \left( \left\langle E^2 \right\rangle - \left\langle E \right\rangle^2 \right)$, where $E$ is the total energy of the system.
The order parameter is calculated as an average over all unit cells, each containing four magnetic sites, i.e. $C = \left| \left\langle S_1 + S_2 - S_3 - S_4 \right\rangle \right|$. 
The curves shown in Fig. \ref{fig:simulation} are then the average quantities over the 5 configurations.

\bigskip
\noindent
{\bf \large Data availability}
\medskip

\noindent The data used in this study is available from E.F. upon request.

\bigskip
\noindent
{\bf \large Code availability}
\medskip

\noindent The Monte Carlo code used in this study is available from E.F. upon request.


\begin{thebibliography}{10}
\expandafter\ifx\csname url\endcsname\relax
  \def\url#1{\texttt{#1}}\fi
\expandafter\ifx\csname urlprefix\endcsname\relax\def\urlprefix{URL }\fi
\providecommand{\bibinfo}[2]{#2}
\providecommand{\eprint}[2][]{\url{#2}}

\bibitem{spaldin2019}
\bibinfo{author}{Spaldin, N.~A.} \& \bibinfo{author}{Ramesh, R.}
\newblock \bibinfo{title}{Advances in magnetoelectric multiferroics}.
\newblock \emph{\bibinfo{journal}{Nature Materials}}
  \textbf{\bibinfo{volume}{18}}, \bibinfo{pages}{203--212}
  (\bibinfo{year}{2019}).
\newblock \urlprefix\url{https://doi.org/10.1038/s41563-018-0275-2}.

\bibitem{liang2021}
\bibinfo{author}{Liang, X.}, \bibinfo{author}{Chen, H.} \&
  \bibinfo{author}{Sun, N.~X.}
\newblock \bibinfo{title}{Magnetoelectric materials and devices}.
\newblock \emph{\bibinfo{journal}{APL Mater.}} \textbf{\bibinfo{volume}{9}},
  \bibinfo{pages}{041114} (\bibinfo{year}{2021}).
\newblock \urlprefix\url{https://doi.org/10.1063/5.0044532}.

\bibitem{pradhan2020}
\bibinfo{author}{Pradhan, D.~K.}, \bibinfo{author}{Kumari, S.} \&
  \bibinfo{author}{Rack, P.~D.}
\newblock \bibinfo{title}{Magnetoelectric composites: Applications, coupling
  mechanisms, and future directions}.
\newblock \emph{\bibinfo{journal}{Nanomaterials}}
  \textbf{\bibinfo{volume}{10}}, \bibinfo{pages}{2072} (\bibinfo{year}{2020}).

\bibitem{Skyrmion}
\bibinfo{author}{Ba, Y.} \emph{et~al.}
\newblock \bibinfo{title}{Electric-field control of skyrmions in multiferroic
  heterostructure via magnetoelectric coupling}.
\newblock \emph{\bibinfo{journal}{Nature Communications}}
  \textbf{\bibinfo{volume}{12}}, \bibinfo{pages}{322} (\bibinfo{year}{2021}).

\bibitem{manipatruni2019}
\bibinfo{author}{Manipatruni, S.} \emph{et~al.}
\newblock \bibinfo{title}{Scalable energy-efficient magnetoelectric spin-orbit
  logic}.
\newblock \emph{\bibinfo{journal}{Nature}} \textbf{\bibinfo{volume}{565}},
  \bibinfo{pages}{35--42} (\bibinfo{year}{2019}).
\newblock \urlprefix\url{http://dx.doi.org/10.1038/s41586-018-0770-2}.

\bibitem{trier2022}
\bibinfo{author}{Trier, F.} \emph{et~al.}
\newblock \bibinfo{title}{Oxide spin-orbitronics: spin- charge interconvension
  and topological textures}.
\newblock \emph{\bibinfo{journal}{Nature Reviews Materials}}
  \textbf{\bibinfo{volume}{7}}, \bibinfo{pages}{258--274}
  (\bibinfo{year}{(2022)}).

\bibitem{nan2017}
\bibinfo{author}{Nan, T.} \emph{et~al.}
\newblock \bibinfo{title}{Acoustically actuated ultra-compact nems
  magnetoelectric antennas}.
\newblock \emph{\bibinfo{journal}{Nature Comms.}} \textbf{\bibinfo{volume}{8}},
  \bibinfo{pages}{296} (\bibinfo{year}{2017}).
\newblock \urlprefix\url{https://doi.org/10.1038/s41467-017-00343-8}.

\bibitem{gu2019}
\bibinfo{author}{Gu, Y.} \emph{et~al.}
\newblock \bibinfo{title}{Mini review on flexible and wearable electronics for
  monitoring human health information}.
\newblock \emph{\bibinfo{journal}{Nanoscale Researsch Letters}}
  \textbf{\bibinfo{volume}{14}}, \bibinfo{pages}{263} (\bibinfo{year}{2019}).
\newblock \urlprefix\url{https://doi.org/10.1186/s11671-019-3084-x}.

\bibitem{dong2020}
\bibinfo{author}{Dong, C.} \emph{et~al.}
\newblock \bibinfo{title}{A portable very low frequency (vlf) communication
  system based on acoustically actuated magnetoelectric antennas}.
\newblock \emph{\bibinfo{journal}{IEEE Antennas and wireless propagation
  letters}} \textbf{\bibinfo{volume}{19}}, \bibinfo{pages}{398--402}
  (\bibinfo{year}{2020}).
\newblock \urlprefix\url{https://doi.org/10.1109/LAWP.2020.2968604}.

\bibitem{hu2011}
\bibinfo{author}{Hu, J.-M.}, \bibinfo{author}{Li, Z.}, \bibinfo{author}{Chen,
  L.-Q.} \& \bibinfo{author}{Nan, C.-W.}
\newblock \bibinfo{title}{High-density magnetoresistive random access memory
  operating at ultralow voltage at room temperature}.
\newblock \emph{\bibinfo{journal}{Nature Comms.}} \textbf{\bibinfo{volume}{2}},
  \bibinfo{pages}{553} (\bibinfo{year}{2011}).
\newblock \urlprefix\url{http://dx.doi.org/10.1038/ncomms1564}.

\bibitem{liu2013}
\bibinfo{author}{Liu, M.} \emph{et~al.}
\newblock \bibinfo{title}{Voltage tuning of ferromagnetic resonance with
  bistable magnetization switching in energy-efficient magnetoelectric
  composites}.
\newblock \emph{\bibinfo{journal}{Advanced Materials}}
  \textbf{\bibinfo{volume}{25}}, \bibinfo{pages}{1435–1439}
  (\bibinfo{year}{2013}).
\newblock \urlprefix\url{http://dx.doi.org/10.1002/adma.201203792}.

\bibitem{matsukura2015}
\bibinfo{author}{Matsukura, F.}, \bibinfo{author}{Tokura, Y.} \&
  \bibinfo{author}{Ohno, H.}
\newblock \bibinfo{title}{Control of magnetism by electric fields}.
\newblock \emph{\bibinfo{journal}{Nature Nanotechnology}}
  \textbf{\bibinfo{volume}{10}}, \bibinfo{pages}{209--220}
  (\bibinfo{year}{2015}).
\newblock \urlprefix\url{http://dx.doi.org/10.1038/NNANO.2015.22}.

\bibitem{eerenstein2006}
\bibinfo{author}{Eerenstein, W.}, \bibinfo{author}{Mathur, N.~D.} \&
  \bibinfo{author}{Scott, J.~F.}
\newblock \bibinfo{title}{Multiferroic and magnetoelectric materials}.
\newblock \emph{\bibinfo{journal}{Nature}} \textbf{\bibinfo{volume}{442}},
  \bibinfo{pages}{759--765} (\bibinfo{year}{2006}).
\newblock \urlprefix\url{https://doi.org/10.1038/nature05023}.

\bibitem{rivera2009}
\bibinfo{author}{Rivera, J.-P.}
\newblock \bibinfo{title}{A short review of the magnetoelectric effect and
  related experimental techniques on single phase (multi-) ferroics}.
\newblock \emph{\bibinfo{journal}{Eur. Phys. J. B}}
  \textbf{\bibinfo{volume}{71}}, \bibinfo{pages}{299–313}
  (\bibinfo{year}{2009}).
\newblock \urlprefix\url{http://dx.doi.org/10.1140/epjb/e2009-00336-7}.

\bibitem{schmid2008}
\bibinfo{author}{Schmid, H.}
\newblock \bibinfo{title}{Some symmetry aspects of ferroics and single phase
  multiferroics}.
\newblock \emph{\bibinfo{journal}{Journal of Physics: Condensed Matter, Volume
  20, Number 43}} \textbf{\bibinfo{volume}{20}}, \bibinfo{pages}{43}
  (\bibinfo{year}{2008}).
\newblock \urlprefix\url{http://dx.doi.org/10.1088/0953-8984/20/43/434201}.

\bibitem{fishman1978}
\bibinfo{author}{Fishman, S.} \& \bibinfo{author}{Aharony, A.}
\newblock \bibinfo{title}{Phase diagrams of multicritical points in randomly
  mixed magnets. i. mixed anisotropies}.
\newblock \emph{\bibinfo{journal}{Phys. Rev. B}} \textbf{\bibinfo{volume}{18}},
  \bibinfo{pages}{3507--3520} (\bibinfo{year}{1978}).
\newblock \urlprefix\url{https://doi.org/10.1103/PhysRevB.18.3507}.

\bibitem{oguchi1978}
\bibinfo{author}{Oguchi, T.} \& \bibinfo{author}{Ishikawa, T.}
\newblock \bibinfo{title}{Theory of a mixture of two
  anisotropicantiferromagnets with different easy axis}.
\newblock \emph{\bibinfo{journal}{Journal of the Physical Society of Japan}}
  \textbf{\bibinfo{volume}{45}}, \bibinfo{pages}{1213} (\bibinfo{year}{1978}).
\newblock \urlprefix\url{https://doi.org/10.1143/JPSJ.45.1213}.

\bibitem{mano1990}
\bibinfo{author}{Mano, H.}
\newblock \bibinfo{title}{Possible phase diagrams of systems with compteting
  anisotropies}.
\newblock \emph{\bibinfo{journal}{Prog. Theo. Phys. Suppl.}}
  \textbf{\bibinfo{volume}{101}}, \bibinfo{pages}{597} (\bibinfo{year}{1990}).
\newblock \urlprefix\url{https://doi.org/10.1143/PTP.101.597}.

\bibitem{vaknin2004}
\bibinfo{author}{Vaknin, D.}, \bibinfo{author}{Zarestky, J.~L.},
  \bibinfo{author}{Rivera, J.-P.} \& \bibinfo{author}{Schmid, H.}
\newblock \bibinfo{title}{Commensurate-incommensurate magnetic phase transition
  in magnetoelectric single crystal {LiNiPO$_4$}}.
\newblock \emph{\bibinfo{journal}{Phys. Rev. Lett.}}
  \textbf{\bibinfo{volume}{92}}, \bibinfo{pages}{207201}
  (\bibinfo{year}{2004}).
\newblock \urlprefix\url{https://doi.org/10.1103/PhysRevLett.92.207201}.

\bibitem{toftpetersen2011}
\bibinfo{author}{Toft-Petersen, R.} \emph{et~al.}
\newblock \bibinfo{title}{High-field magnetic phase transitions and spin
  excitations in magnetoelectric {LiNiPO$_4$}}.
\newblock \emph{\bibinfo{journal}{Phys. Rev. B}} \textbf{\bibinfo{volume}{84}},
  \bibinfo{pages}{054408} (\bibinfo{year}{2011}).
\newblock \urlprefix\url{https://doi.org/10.1103/PhysRevB.84.054408}.

\bibitem{liang2008}
\bibinfo{author}{Liang, G.} \emph{et~al.}
\newblock \bibinfo{title}{Anisotropy in magnetic properties and electronic
  structure of single-crystal {LiFePO$_4$}}.
\newblock \emph{\bibinfo{journal}{Phys. Rev. B}} \textbf{\bibinfo{volume}{77}},
  \bibinfo{pages}{064414} (\bibinfo{year}{2008}).
\newblock \urlprefix\url{http://dx.doi.org/10.1103/PhysRevB.77.064414}.

\bibitem{MercierThesis}
\bibinfo{author}{Mercier, M.}
\newblock \bibinfo{title}{Ph.d. \ thesis}.
\newblock \emph{\bibinfo{journal}{Universit\'e de Grenoble}}
  (\bibinfo{year}{1969}).

\bibitem{santoro1966}
\bibinfo{author}{Santoro, R.~P.}, \bibinfo{author}{Segal, D.~J.} \&
  \bibinfo{author}{Newnham, R.~E.}
\newblock \bibinfo{title}{Magnetic properties of {LiCoPO$_4$} and
  {LiNiPO$_4$}}.
\newblock \emph{\bibinfo{journal}{J. Phys. Chem. Solids.}}
  \textbf{\bibinfo{volume}{27}}, \bibinfo{pages}{1192--1193}
  (\bibinfo{year}{1966}).
\newblock \urlprefix\url{https://doi.org/10.1016/0022-3697(66)90097-7}.

\bibitem{jensen2009_struc}
\bibinfo{author}{Jensen, T. B.~S.} \emph{et~al.}
\newblock \bibinfo{title}{Field-induced magnetic phases and electric
  polarization in {LiNiPO$_4$}}.
\newblock \emph{\bibinfo{journal}{Phys. Rev. B}} \textbf{\bibinfo{volume}{79}},
  \bibinfo{pages}{092413} (\bibinfo{year}{2009}).
\newblock \urlprefix\url{https://doi.org/10.1103/PhysRevB.79.092412}.

\bibitem{santoro1967}
\bibinfo{author}{Santoro, R.~P.} \& \bibinfo{author}{Newnham, R.~E.}
\newblock \bibinfo{title}{Antiferromagnetism in {LiFePO$_4$}}.
\newblock \emph{\bibinfo{journal}{Acta Cryst.}} \textbf{\bibinfo{volume}{22}},
  \bibinfo{pages}{344--347} (\bibinfo{year}{1967}).
\newblock \urlprefix\url{https://doi.org/10.1107/S0365110X67000672}.

\bibitem{li2006}
\bibinfo{author}{Li, J.}, \bibinfo{author}{Garlea, V.~O.},
  \bibinfo{author}{Zarestky, J.~L.} \& \bibinfo{author}{Vaknin, D.}
\newblock \bibinfo{title}{Spin-waves in antiferromagnetic single-crystal
  {LiFePO$_4$}}.
\newblock \emph{\bibinfo{journal}{Phys. Rev. B}} \textbf{\bibinfo{volume}{73}},
  \bibinfo{pages}{024410} (\bibinfo{year}{2006}).
\newblock \urlprefix\url{https://doi.org/10.1103/PhysRevB.73.024410}.

\bibitem{jensen2009_dyna}
\bibinfo{author}{Jensen, T. B.~S.} \emph{et~al.}
\newblock \bibinfo{title}{Anomalous spin waves and the
  commensurate-incommensurate magnetic phase transition in {LiNiPO$_4$}}.
\newblock \emph{\bibinfo{journal}{Phys. Rev. B}} \textbf{\bibinfo{volume}{79}},
  \bibinfo{pages}{092413} (\bibinfo{year}{2009}).
\newblock \urlprefix\url{https://doi.org/10.1103/PhysRevB.79.092413}.

\bibitem{toftpetersen2015}
\bibinfo{author}{Toft-Petersen, R.} \emph{et~al.}
\newblock \bibinfo{title}{Anomalous magnetic structure and spin dynamics in
  magnetoelectric {LiFePO$_4$}}.
\newblock \emph{\bibinfo{journal}{Phys. Rev. B}} \textbf{\bibinfo{volume}{92}},
  \bibinfo{pages}{024404} (\bibinfo{year}{2015}).
\newblock \urlprefix\url{https://doi.org/10.1103/PhysRevB.92.024404}.

\bibitem{zimmermann2013}
\bibinfo{author}{Zimmermann, A.~S.}, \bibinfo{author}{Sondermann, E.},
  \bibinfo{author}{Li, J.}, \bibinfo{author}{Vaknin, D.} \&
  \bibinfo{author}{Fiebig, M.}
\newblock \bibinfo{title}{Antiferromagnetic order in
  {Li(Ni$_{1-x}$Fe$_x$)PO$_4$} ($x$ = 0.06, 0.20)}.
\newblock \emph{\bibinfo{journal}{Phys. Rev. B}} \textbf{\bibinfo{volume}{88}},
  \bibinfo{pages}{014420} (\bibinfo{year}{2013}).
\newblock \urlprefix\url{https://doi.org/10.1103/PhysRevB.88.014420}.

\bibitem{li2009}
\bibinfo{author}{Li, J.} \emph{et~al.}
\newblock \bibinfo{title}{Tweaking the spin-wave dispersion and suppressing the
  incommensurate phase in {LiNiPO$_4$} by iron substitution}.
\newblock \emph{\bibinfo{journal}{Phys. Rev. B}} \textbf{\bibinfo{volume}{79}},
  \bibinfo{pages}{174435} (\bibinfo{year}{2009}).
\newblock \urlprefix\url{https://doi.org/10.1103/PhysRevB.79.174435}.

\bibitem{stewart2009}
\bibinfo{author}{Stewart, J.~R.} \emph{et~al.}
\newblock \bibinfo{title}{Disordered materials studied using neutron
  polarization analysis on the multi-detector spectrometer, {D7}}.
\newblock \emph{\bibinfo{journal}{J. Appl. Cryst.}}
  \textbf{\bibinfo{volume}{42}}, \bibinfo{pages}{69--84}
  (\bibinfo{year}{2009}).
\newblock \urlprefix\url{https://doi.org/10.1107/S0021889808039162}.

\bibitem{werner2019}
\bibinfo{author}{Werner, J.} \emph{et~al.}
\newblock \bibinfo{title}{High magnetic field phase diagram and failure of the
  magnetic {Grüneisen} scaling in {LiFePO$_4$}}.
\newblock \emph{\bibinfo{journal}{Phys. Rev. B}} \textbf{\bibinfo{volume}{{\bf
  99}}}, \bibinfo{pages}{214432} (\bibinfo{year}{(2019)}).

\bibitem{Aharony_PRB_1979}
\bibinfo{author}{Fishman, S.} \& \bibinfo{author}{Aharony, A.}
\newblock \bibinfo{title}{Phase diagrams and multicritical points in randomly
  mixed magnets.ii. ferromagnet-antiferromagnet alloys}.
\newblock \emph{\bibinfo{journal}{Phys. Rev. B}} \textbf{\bibinfo{volume}{19}},
  \bibinfo{pages}{3776} (\bibinfo{year}{1979}).

\bibitem{fogh2022_Fe}
\bibinfo{author}{Fogh, E.} \emph{et~al.}
\newblock \bibinfo{title}{The magnetoelectric effect in {LiFePO$_4$} --
  revisited}.
\newblock \bibinfo{note}{Https://ssrn.com/abstract=4157288}.

\bibitem{mercier1968}
\bibinfo{author}{Mercier, M.}, \bibinfo{author}{Bauer, P.} \&
  \bibinfo{author}{Fouilleux, B.}
\newblock \bibinfo{title}{Mesures magnétoélectriques sur {LiFePO$_4$}}.
\newblock \emph{\bibinfo{journal}{C. R. Acad. Sc. Paris B}}
  \textbf{\bibinfo{volume}{267}}, \bibinfo{pages}{1345--1346}
  (\bibinfo{year}{1968}).

\bibitem{aharony1975}
\bibinfo{author}{Aharony, A.}
\newblock \bibinfo{title}{Tetracritical points in mixed magnetic crystals}.
\newblock \emph{\bibinfo{journal}{Phys. Rev. Lett}}
  \textbf{\bibinfo{volume}{34}}, \bibinfo{pages}{590} (\bibinfo{year}{1975}).
\newblock \urlprefix\url{https://doi.org/10.1103/PhysRevLett.34.590}.

\bibitem{lindgaard1976}
\bibinfo{author}{Lindgaard, P.-A.}
\newblock \bibinfo{title}{Theory of random anisotropic magnetic alloys}.
\newblock \emph{\bibinfo{journal}{Phys. Rev. B}} \textbf{\bibinfo{volume}{14}},
  \bibinfo{pages}{4074} (\bibinfo{year}{1976}).
\newblock \urlprefix\url{https://doi.org/10.1063/1.30380}.

\bibitem{matsubara1977}
\bibinfo{author}{Matsubara, F.} \& \bibinfo{author}{Inawashiro, S.}
\newblock \bibinfo{title}{Mixture of two anisotropic antiferromagnets with
  different easy axis}.
\newblock \emph{\bibinfo{journal}{Journal of the Physical Society of Japan}}
  \textbf{\bibinfo{volume}{42}}, \bibinfo{pages}{1529} (\bibinfo{year}{1977}).
\newblock \urlprefix\url{https://doi.org/10.1143/JPSJ.42.1529}.

\bibitem{mano1985}
\bibinfo{author}{Mano, H.}
\newblock \bibinfo{title}{Pair spin theory of a random micture with competing
  ising and xy anisotropies}.
\newblock \emph{\bibinfo{journal}{Journal of the Physical Society of Japan}}
  \textbf{\bibinfo{volume}{55}}, \bibinfo{pages}{908--919}
  (\bibinfo{year}{1986}).
\newblock \urlprefix\url{https://doi.org/10.1143/JPSJ.55.908}.

\bibitem{wong1980}
\bibinfo{author}{Wong, P.}, \bibinfo{author}{Horn, P.~M.},
  \bibinfo{author}{Birgeneau, H.~J.}, \bibinfo{author}{Safinya, C.~H.} \&
  \bibinfo{author}{Shirane, G.}
\newblock \bibinfo{title}{Competing order parameters in quenched random alloys:
  {Fe$_{1-x}$Co$_x$Cl$_2$}}.
\newblock \emph{\bibinfo{journal}{Phys. Rev. Lett.}}
  \textbf{\bibinfo{volume}{45}}, \bibinfo{pages}{1974--1977}
  (\bibinfo{year}{1980}).
\newblock \urlprefix\url{https://doi.org/10.1103/PhysRevLett.45.1974}.

\bibitem{katsumata1979}
\bibinfo{author}{Katsumata, K.}, \bibinfo{author}{Kobayashi, M.},
  \bibinfo{author}{Sato, T.} \& \bibinfo{author}{Miyako, Y.}
\newblock \bibinfo{title}{Experimental phase diagram of a random mixture of two
  anisotropic antiferromagnets}.
\newblock \emph{\bibinfo{journal}{Phys. Rev. B}} \textbf{\bibinfo{volume}{19}},
  \bibinfo{pages}{2700--2703} (\bibinfo{year}{1979}).
\newblock \urlprefix\url{https://doi.org/10.1103/PhysRevB.19.2700}.

\bibitem{tawaraya1980}
\bibinfo{author}{Tawaraya, T.}, \bibinfo{author}{Katsumata, K.} \&
  \bibinfo{author}{Yoshizawa, H.}
\newblock \bibinfo{title}{Neutron diffraction experiment on a randomly mixed
  antiferromagnet with competing spin anisotropies}.
\newblock \emph{\bibinfo{journal}{J. Phys. Soc. Jpn.}}
  \textbf{\bibinfo{volume}{49}}, \bibinfo{pages}{1299--1305}
  (\bibinfo{year}{1980}).
\newblock \urlprefix\url{https://doi.org/10.1143/JPSJ.49.1299}.

\bibitem{perez2015}
\bibinfo{author}{Perez, F.~A.} \emph{et~al.}
\newblock \bibinfo{title}{Phase diagram of a three-dimensional antiferromagnet
  with random magnetic anisotropy}.
\newblock \emph{\bibinfo{journal}{Phys. Rev. Lett.}}
  \textbf{\bibinfo{volume}{114}}, \bibinfo{pages}{097201}
  (\bibinfo{year}{2015}).
\newblock \urlprefix\url{https://doi.org/10.1103/PhysRevLett.114.097201}.

\bibitem{bevaart1978}
\bibinfo{author}{Bevaart, L.}, \bibinfo{author}{Frikkee, E.},
  \bibinfo{author}{Lebesque, J.} \& \bibinfo{author}{de~Jongh, L.~J.}
\newblock \bibinfo{title}{Magnetic and neutron scattering experiments on the
  antiferromagnetic layer-type compounds {K$_2$Mn$_{1-x}M_x$F$_4$ ($M$ =
  Fe,Co)}}.
\newblock \emph{\bibinfo{journal}{Phys. Rev. B}} \textbf{\bibinfo{volume}{18}},
  \bibinfo{pages}{3376--3392} (\bibinfo{year}{1978}).
\newblock \urlprefix\url{https://doi.org/10.1103/PhysRevB.18.3376}.

\bibitem{wegner1973}
\bibinfo{author}{Wegner, F.}
\newblock \bibinfo{title}{On the magnetic phase diagram of {(Mn,Fe)WO$_4$}}.
\newblock \emph{\bibinfo{journal}{Solid State Comm.}}
  \textbf{\bibinfo{volume}{12}}, \bibinfo{pages}{785--787}
  (\bibinfo{year}{1973}).
\newblock \urlprefix\url{https://doi.org/10.1016/0038-1098(73)90839-9}.

\bibitem{toftpetersen2017}
\bibinfo{author}{Toft-Petersen, R.} \emph{et~al.}
\newblock \bibinfo{title}{Field-induced reentrant magnetoelectric phase in
  {LiNiPO$_4$}}.
\newblock \emph{\bibinfo{journal}{Phys. Rev. B}} \textbf{\bibinfo{volume}{95}},
  \bibinfo{pages}{064421} (\bibinfo{year}{2017}).
\newblock \urlprefix\url{https://doi.org/10.1103/PhysRevB.95.064421}.

\bibitem{fogh2020}
\bibinfo{author}{Fogh, E.} \emph{et~al.}
\newblock \bibinfo{title}{Magnetic structures and quadratic magnetoelectric
  effect in {LiNiPO$_4$} beyond {30T}}.
\newblock \emph{\bibinfo{journal}{Phys. Rev. B}}
  \textbf{\bibinfo{volume}{101}}, \bibinfo{pages}{024403}
  (\bibinfo{year}{2020}).
\newblock \urlprefix\url{https://doi.org/10.1103/PhysRevB.101.024403}.

\bibitem{diethrich2022}
\bibinfo{author}{Diethrich, T.~J.}, \bibinfo{author}{Gnewuch, S.},
  \bibinfo{author}{Dold, K.~G.}, \bibinfo{author}{Taddei, K.~M.} \&
  \bibinfo{author}{Rodriguez, E.~E.}
\newblock \bibinfo{title}{Tuning magnetic symmetry and properties in the
  olivine series {Li$_{1-x}$Fe$_x$Mn$_{1-x}$PO$_4$} through selective
  delithiation}.
\newblock \emph{\bibinfo{journal}{Chem. Mater.}} \textbf{\bibinfo{volume}{{\bf
  34}}}, \bibinfo{pages}{5039--5053} (\bibinfo{year}{(2022)}).

\bibitem{semkin2022}
\bibinfo{author}{Semkin, M.~A.} \emph{et~al.}
\newblock \bibinfo{title}{Magnetic phase transitions in the
  {LiNi$_{0.9}M_{0.1}$PO$_4$} {($M$ = Mn, Co)} single crystals}.
\newblock \emph{\bibinfo{journal}{Phys. Scr.}} \textbf{\bibinfo{volume}{{\bf
  97}}}, \bibinfo{pages}{025707} (\bibinfo{year}{(2022)}).

\bibitem{ye2008}
\bibinfo{author}{Ye, F.} \emph{et~al.}
\newblock \bibinfo{title}{Magnetic switching and phase competition in the
  multiferroic antiferromagnet {Mn$_{1-x}$Fe$_x$WO$_4$}}.
\newblock \emph{\bibinfo{journal}{Phys. Rev. B}} \textbf{\bibinfo{volume}{78}},
  \bibinfo{pages}{193101} (\bibinfo{year}{(2008)}).

\bibitem{ye2012}
\bibinfo{author}{Ye, F.} \emph{et~al.}
\newblock \bibinfo{title}{Magnetic order and spin-flop transitions in the
  cobalt-doped multiferroic {Mn$_{1-x}$Co$_x$WO$_4$}}.
\newblock \emph{\bibinfo{journal}{Phys. Rev. B}} \textbf{\bibinfo{volume}{86}},
  \bibinfo{pages}{094429} (\bibinfo{year}{(2012)}).

\bibitem{patureau2015}
\bibinfo{author}{Patureau, P.} \emph{et~al.}
\newblock \bibinfo{title}{{Incorporation of Jahn-Teller Cu$^{2+}$ Ions into
  Magnetoelectric Multiferroic MnWO$_4$: Structural, Magnetic, and Dielectric
  Permittivity Properties of Mn$_{1-x}$Cu$_x$WO$_4$ $(x \leq 0.25)$}}.
\newblock \emph{\bibinfo{journal}{Inorg. Chem.}} \textbf{\bibinfo{volume}{54}},
  \bibinfo{pages}{10623--10631} (\bibinfo{year}{(2015)}).

\bibitem{chaudhury2009}
\bibinfo{author}{Chaudhury, R.~P.}, \bibinfo{author}{Lorenz, B.},
  \bibinfo{author}{Wang, Y.~Q.}, \bibinfo{author}{Sun, Y.~Y.} \&
  \bibinfo{author}{Chu, C.~W.}
\newblock \bibinfo{title}{Re-entrant ferroelectricity and the multiferroic
  phase diagram of {Mn$_{1-x}$Fe$_x$WO$_4$}}.
\newblock \emph{\bibinfo{journal}{New J. Phys.}} \textbf{\bibinfo{volume}{11}},
  \bibinfo{pages}{033036} (\bibinfo{year}{2009}).
\newblock \urlprefix\url{https://doi.org/10.1088/1367-2630/11/3/033036}.

\bibitem{chynoweth1956}
\bibinfo{author}{Chynoweth, A.~G.}
\newblock \bibinfo{title}{Dynamic method for measuring the pyroelectric effect
  with special reference to barium titanate}.
\newblock \emph{\bibinfo{journal}{J. Appl. Phys.}}
  \textbf{\bibinfo{volume}{27}}, \bibinfo{pages}{78--84}
  (\bibinfo{year}{1956}).
\newblock \urlprefix\url{https://doi.org/10.1063/1.1722201}.

\bibitem{paeckelBachelor}
\bibinfo{author}{Paeckel, S.}
\newblock \emph{\bibinfo{title}{Setup of a measurement system to determine the
  electric polarization}}.
\newblock \bibinfo{type}{Bachelor's thesis}, \bibinfo{school}{Technische
  Universität Berlin} (\bibinfo{year}{2010}).

\bibitem{metropolis1953}
\bibinfo{author}{Metropolis, N.}, \bibinfo{author}{Rosenbluth, A.~W.},
  \bibinfo{author}{Rosenbluth, M.~N.} \& \bibinfo{author}{Teller, A.~H.}
\newblock \bibinfo{title}{Equation of state calculations by fast computing
  machines}.
\newblock \emph{\bibinfo{journal}{J. Chem. Phys.}}
  \textbf{\bibinfo{volume}{21}}, \bibinfo{pages}{1087} (\bibinfo{year}{1953}).
\newblock \urlprefix\url{https://doi.org/10.1063/1.1699114}.

\bibitem{abrahams1993}
\bibinfo{author}{Abrahams, I.} \& \bibinfo{author}{Easson, K.~S.}
\newblock \bibinfo{title}{Structure of lithium nickel phosphate}.
\newblock \emph{\bibinfo{journal}{Acta Cryst. C}} \textbf{\bibinfo{volume}{{\bf
  49}}}, \bibinfo{pages}{925--926} (\bibinfo{year}{(1993)}).

\bibitem{moon1969}
\bibinfo{author}{Moon, R.~M.}, \bibinfo{author}{Riste, T.} \&
  \bibinfo{author}{Koehler, W.~C.}
\newblock \bibinfo{title}{Polarization analysis of thermal neutron scattering}.
\newblock \emph{\bibinfo{journal}{Phys. Rev.}} \textbf{\bibinfo{volume}{181}},
  \bibinfo{pages}{920--931} (\bibinfo{year}{1969}).
\newblock \urlprefix\url{https://doi.org/10.1103/PhysRev.181.920}.

\bibitem{ehlers2013}
\bibinfo{author}{Ehlers, G.}, \bibinfo{author}{Stewart, J.~R.},
  \bibinfo{author}{Wildes, A.~R.}, \bibinfo{author}{Deen, P.~P.} \&
  \bibinfo{author}{Andersen, K.~H.}
\newblock \bibinfo{title}{Generalization of the classical xyz-polarization
  analysis technique to out-of-plane and inelastic scattering}.
\newblock \emph{\bibinfo{journal}{Rev. Sci. Instrum.}}
  \textbf{\bibinfo{volume}{84}}, \bibinfo{pages}{093901}
  (\bibinfo{year}{2013}).
\newblock \urlprefix\url{https://doi.org/10.1063/1.4819739}.

\setcounter{firstbib}{\value{NAT@ctr}}

\end{thebibliography}

\bigskip
\noindent


\begin{thebibliography}{1}

\setcounter{NAT@ctr}{\value{firstbib}}

  
  \bibitem{mays1963}
\bibinfo{author}{Mays, J. M.}
\newblock \bibinfo{title}{Nuclear Magnetic Resonances and {Mn-O-P-O-Mn} Superexchange Linkages in Paramagnetic and Antiferromagnetic {LiMnPO$_4$}}.
\newblock \emph{\bibinfo{journal}{Phys. Rev.}} \textbf{\bibinfo{volume}{{\bf
  131}}}, \bibinfo{pages}{38} (\bibinfo{year}{(1963)}).


  \bibitem{rousse2003}
\bibinfo{author}{Rousse, G.}, \bibinfo{author}{Rodriguez-Carvajal, J.},
  \bibinfo{author}{Patoux, S.} \&
  \bibinfo{author}{Masquelier, C.}
\newblock \bibinfo{title}{Magnetic Structures of the Triphylite {LiFePO$_4$} and of Its Delithiated Form {FePO$_4$}}.
\newblock \emph{\bibinfo{journal}{Chem. Mater.}} \textbf{\bibinfo{volume}{{\bf
  15}}}, \bibinfo{pages}{4082--4090} (\bibinfo{year}{(2003)}).



\end{thebibliography}

\bigskip
\noindent
{\bf Acknowledgements} This work was supported by the European Research Council through the Synergy network HERO (Grant No. 810451).
We thank the EU Intereg program MAXESS4FUN for Cross Border and Society for funding the simulation work.
We are grateful for neutron beamtime received for this project at the instruments TriCS and RITA-II at the SINQ neutron spallation source at the Paul Scherrer Institute, at the E5 diffractometer at the BER-II research reactor at the Helmholtz-Zentrum Berlin, and at the 4F1 spectrometer at the research reactor at the Laboratoire Leon Brillouin. 
We acknowledge Diamond Light Source for beamtime on I16.
We are grateful for travel support from the Danish Agency for Science, Technology and Innovation under DANSCATT.
Ames Laboratory is operated for the U.S. Department of Energy by Iowa State University under Contract No. DEAC02-07CH11358.
We thank M. Laver for assistance with neutron scattering experiments, A. Sokolowski for support with pyrocurrent measurements and J. Li for samples.

\bigskip
\noindent
{\bf Author contributions} The experimental project and theoretical framework was conceived by E.F, R.T.-P. and N.B.C.
The crystals were grown by D.V.
Magnetic susceptibility measurements were performed by E.F., K.S.P. and K.K.M. 
Diffraction experiments were performed by E.F., R.E.H., A.B.K., M.K.S., G.M., A.B., J.-R.S. M.R., P.B., S.H.-D., O.Z., J.S., Ch.N. R.T.-P. and N.B.C.   
Pyrocurrent measurements were carried out by E.F., B.K., S.P. and R.T.-P. 
Sample preparation for the pyrocurrent measurements on the parent compounds was performed by E.F. and A.P. 
E.F. implemented and performed Monte Carlo simulations with support from O.F.S. 
Data analysis and figure preparation were performed by E.F. 
Interpretation and manuscript writing was done by E.F., R.T.-P., N.B.C. and H.M.R. with input from all authors.

\bigskip
\noindent
{\bf Competing Interests} The authors declare no competing interests. 

\bigskip
\noindent {\bf Correspondence} and requests for materials should be addressed to E.F., N.B.C. or R.T.-P.


\clearpage

\setcounter{figure}{0}
\renewcommand{\thefigure}{ED\arabic{figure}}

\onecolumngrid

\noindent
{\large {\bf {Supplementary information for}}}
\vskip4mm
\noindent
{\large {\bf {Tuning magnetoelectricity in a mixed-anisotropy antiferromagnet}}}
\vskip4mm

\noindent
Ellen Fogh, Bastian Klemke, Manfred Reehuis, Philippe Bourges, Christof Niedermayer, 
Sonja Holm-Dahlin, 
Oksana Zaharko, Jürg Schefer, Andreas B. Kristensen, Michael K. Sørensen, Sebastian Paeckel, Kasper S. Pedersen, 
Rasmus E. Hansen, Alexandre Pages, 
Kimmie K. Moerner, Giulia Meucci, Jian-Rui Soh, 
Alessandro Bombardi, David Vaknin, 
Henrik M. Rønnow, 
Olav F. Syljuåsen. Niels B. Christensen and Rasmus Toft-Petersen

\vskip6mm

\twocolumngrid

\begin{table*}[t!]
    \caption{Structure factors for various magnetic Bragg peaks as composed by the four basis vectors $A = \left( \uparrow \downarrow \downarrow \uparrow \right)$, $G = \left( \uparrow \downarrow \uparrow \downarrow \right)$, $C = \left( \uparrow \uparrow \downarrow \downarrow \right)$ and $F = \left( \uparrow \uparrow \uparrow \uparrow \right)$. The last three columns contain the factors $\mid\!\! \mathcal{P}_\alpha({\bf Q})\!\!\mid^2$ for the three crystallographic axes, $\alpha = \lbrace a,b,c \rbrace$. The factors are normalized to unit spin length.}
    \label{tab:structurefactors}
    \centering
    \begin{tabular}{| c | c c c c | c c c |}
        \hline
        ${\bf Q}$   & $\mid\!\! S_A(\bf Q) \!\!\mid^2$   & $\mid\!\! S_G(\bf Q) \!\!\mid^2$   & $\mid\!\! S_C(\bf Q) \!\!\mid^2$   & $\mid\!\! S_F(\bf Q) \!\!\mid^2$   & $\mid\!\! \mathcal{P}_a(\bf Q) \!\!\mid^2$   & $\mid\!\! \mathcal{P}_b(\bf Q) \!\!\mid^2$   &$ \mid\!\! \mathcal{P}_c(\bf Q) \!\!\mid^2$ \\
        \hline
        $(0,0,\pm 1)$  & 0     & 15.05 & 0.50  & 0     & 1     & 1     & 0\\
        $(0,1,0)$   & 0     & 0     & 16    & 0     & 1     & 0     & 1\\
        $(3,0,-1)$  & 0.16  & 0.35  & 10.60 & 4.91  & 0.34  & 1     & 0.66\\
        $(0,1,-2)$   & 0     & 1.93  & 14.07 & 0     & 1     & 0.86  & 0.14\\
        $(1,0,0)$   & 15.37 & 0.63  & 0     & 0     & 0     & 1     & 1\\
        $(1,1,0)$   & 0.63  & 15.37 & 0     & 0     & 0.75  & 0.25  & 1\\
        $(1,2,0)$   & 15.37 & 0.63  & 0     & 0     & 0.92  & 0.08  & 1\\
        $(2,1,0)$   & 0     & 0     & 13.59 & 2.41  & 0.41  & 0.58  & 1\\
        \hline
    \end{tabular}
\end{table*}

\noindent {\large \bf Neutron diffraction experiments}
\medskip

In the crystal structure of Li$M$PO$_4$, the transition metal ions reside on the $4c$ Wyckoff positions with coordinates $r_1=(1+\varepsilon,1/4, 1-\delta)$, $r_2=(3/4+\varepsilon,1/4,1/2+\delta)$, $r_3=(3/4-\varepsilon,1/4,\delta)$ and $r_4=(1/4-\varepsilon,1/4,1/2-\delta)$, with $\varepsilon = 0.025$, $\delta = 0.0175$ for \Ni\ \cite{abrahams1993} and $\varepsilon = 0.032$, $\delta = 0.0252$ for \Fe\ \cite{toftpetersen2015}. To analyse the magnetic structures with propagation vector ${\bf k }=(0,0,0)$ that are experimentally found in Li$M$PO$_4$ \cite{santoro1966,santoro1967,mays1963}, we consider four basis vectors corresponding to the specific sequences of spins $A_\alpha = \left( \uparrow \downarrow \downarrow \uparrow \right)$, $G_\alpha = \left( \uparrow \downarrow \uparrow \downarrow \right)$, $C_\alpha = \left( \uparrow \uparrow \downarrow \downarrow \right)$ and $F_\alpha = \left( \uparrow \uparrow \uparrow \uparrow \right)$ on the sites $r_1$-$r_4$. When combining these four basis vectors with three orthogonal spin directions, the resulting 12 components allow a full description of the possible magnetic structures. 

For unpolarized neutrons the differential scattering cross-section corresponding to each component of the magnetic structure is proportional to $\mid\!\! S_R({\bf Q})\!\!\mid^2\mid\!\!\mathcal{P}({\bf Q})\!\!\mid^2$, where 
\begin{equation*}
S_R ({\bf Q})=\sum_d m^R_d e^{i {\bf Q} \cdot {\bf r}_d},
\;\;\;{\mathcal{P}}({\bf Q})=\hat{{\bf Q}}\times({\hat{\bf e}}\times\hat{{\bf Q}}).
\end{equation*}
Here $S_R({\bf Q})$ is a structure factor involving the summation over the magnetic moments of the ions at sites $r_1$-$r_4$ and $R$ refers to the particular basis vector of the magnetic structure. The vector $\hat{\bf Q}$ is a unit vector along the neutron momentum transfer in the scattering process, and ${\hat{\bf e}}$ is a unit vector along the direction of the magnetic moments for the component $R$. The factor ${\mathcal{P}}(\bf Q)$ reflects the fact that neutrons only scatter from electronic spin components perpendicular to $\bf Q$. 

Table \ref{tab:structurefactors} lists $\mid\!\! S_R(\bf Q) \!\!\mid^2$ and $\mid\!\! \mathcal{P}(\bf Q) \!\!\mid^2$ for the basis vectors $C$, $A$, $G$, $F$ and spin directions ${\hat{\bf e}}$ parallel to $a$, $b$ or $c$ for the Bragg peaks probed in our experiments. The entries of the table represent the sensitivity of each magnetic Bragg peak to the different basis structure components. For example, the peaks $(0,0,\pm 1)$ are dominated by components corresponding to the basis vector $G_a$ and $G_b$ with minor contributions from components $C_a$ and $C_b$. On the other hand, observing $(0,1,0)$ implies that at least one of the components $C_a$ and $C_c$ must be present. If the magnetic structure was exclusively described by the $C_y$ component, the $(0,1,0)$ peak would have no intensity because the magnetic moments are parallel to ${\bf Q}$ and from which follows ${\mathcal{P}}_b({\bf Q})=0$.

For our experiments using uniaxial polarization analysis, the relevant expressions for the non spin-flip (NSF) scattering intensities are 
\begin{eqnarray}
I_x^{\rm NSF} &=& I_{\rm coh}^N + B_{\rm NSF} \nonumber \\
I_y^{\rm NSF} &=& I_{\rm coh}^N + I_{y}^M + B_{\rm NSF} \nonumber \\
I_z^{\rm NSF} &=& I_{\rm coh}^N + I_{z}^M + B_{\rm NSF} \nonumber 
\end{eqnarray}
while those for spin-flip (SF) intensities are
\begin{eqnarray}
I_x^{\rm SF}  &=& I_{y}^M + I_{z}^M + B_{\rm SF} \nonumber \\ 
I_y^{\rm SF}  &=& I_{z}^M + B_{\rm SF} \nonumber \\ 
I_z^{\rm SF}  &=& I_{y}^M + B_{\rm SF} \nonumber. 
\end{eqnarray}
These expressions are valid when contributions from nuclear-magnetic interference scattering and chiral scattering can be neglected \cite{moon1969}. The subscripts $x$, $y$ and $z$ denote to the direction of neutron polarization where $x$ corresponds to the configuration where the neutron polarization is parallel to ${\bf Q}$, $y$ is when the polarization is perpendicular to ${\bf Q}$ and in the horizontal scattering plane, while $z$ refers to when the polarization is perpendicular to the scattering plane. The coordinate system is illustrated in Fig. \ref{fig:4F1scans}c.
The contribution $I_{\rm coh}^N$ to the three NSF cross-sections corresponds to nuclear Bragg scattering. 
The terms $I_{y}^M$ and $I_{z}^M$ are magnetic contributions to a given Bragg peak intensity, arising from electronic spin components parallel to $y$ and $z$, respectively. Note that the NSF contributions carry information on spin components along the neutron polarization, while the SF contributions carry information on spin components perpendicular to the neutron polarization.
The background contribution $B_{\rm NSF}$ in the NSF channels involves the sum of nuclear isotope incoherent scattering and one third of the total nuclear spin incoherent scattering. The background term $B_{\rm SF}$ in the SF channels involves the remaining two thirds of the nuclear spin incoherent scattering.  

Because the polarization of the neutron beam is never perfect, a correction is needed before the measured Bragg peak intensities can be directly compared to the above expressions for the polarized neutron diffraction intensity. The relevant correction factors, also called flipping ratios, were determined from measurements of nuclear Bragg peaks above $T_2$. We found flipping ratios $40$ ($25$) for nuclear scattering vectors along the $b^*$ ($c^*$) direction, respectively. For more details on the analysis see e.g. Refs. \cite{stewart2009}, \cite{moon1969} or \cite{ehlers2013}.

\noindent {\bf E5 experiment.} The room temperature crystal structure of \NiFe\ was determined from 3255 reflections from which 724 are unique. The refinement of a total of 41 parameters gives a residual of $R_F = \sum \frac{|F_{\mathrm{obs}}|-|F_{\mathrm{calc}}|}{|F_{\mathrm{obs}}|} = 0.050$ where $F_{\mathrm{obs}}$ and $F_{\mathrm{calc}}$ are the observed and calculated intensities, respectively. We used the following lattice parameters as obtained from UB matrix calculations using strong reflections during the experiment setup: $a = 9.9741(14)\,\mathrm{Å}$, $b = 5.8284(6)\,\mathrm{Å}$, $c = 4.6326(5)\,\mathrm{Å}$. The refinement results are summarized in Table \ref{tab:E5refinement}. The ion displacements away from a perfect face centered configuration are $\varepsilon = 0.02701(5)$ and $\delta = 0.01877(10)$, values which not surprisingly lie in between those of the parent compounds.

\begin{table*}
    \caption{Results of the crystal structure refinements of \NiFe\ from single-crystal neutron diffraction data ($\lambda = 0.896\,\mathrm{Å}$) collected at room temperature. The refinement of the data set was carried out using space group \textit{Pnma}. The thermal parameters $U_{ij}$ (given in $100\,\mathrm{Å}^2$) are in the form $\exp \left[ -2\pi^2 \left( U_{11} H^2 (a^*)^2 + ... + 2 U_{13} HL (a^*) (c^*) \right) \right]$, where $H$, $K$ and $L$ are Miller indices and $a^*$, $b^*$ and $c^*$ are the reciprocal lattice parameters. For symmetry reasons the values $U_{12}$ and $U_{23}$ of the atoms located at the Wyckoff position $4c$ are zero for $Pnma$. Site occupancies were refined for Li and O sites only. Within the errors only Li sites do not reach full occupancy.}
    \label{tab:E5refinement}
    \centering
    \begin{tabular}{|c c  c c c  c c c c c c  c|}
        \hline
        Atom & Site & $x$ & $y$ & $z$ & $U_{11}$ & $U_{22}$ & $U_{33}$    & $U_{12}$    & $U_{13}$   & $U_{23}$  & Occ.\\
        \hline
        Li  & $4a$  & 0 & 0 & 0 & 1.75(19)  & 1.82(19)  & 1.20(14)  & 0.10(17)  & -0.41(10) & -0.51(10)   & 0.94(3)\\
        Fe  & $4c$  & 0.27701(5) & $1/4$ & 0.98123(10) & 0.36(2) & 0.61(2) & 0.57(2) & 0 & 0.00(1) & 0 & 0.2\\
        Ni  & $4c$ & 0.27701(5) & $1/4$ & 0.98123(10) & 0.36(2) & 0.61(2) & 0.57(2) & 0 & 0.00(1) & 0 & 0.8\\
        P  & $4c$ & 0.09461(10) & $1/4$ & 0.41758(19) & 0.34(4) & 0.59(4) & 0.37(3) & 0 & 0.01(2) & 0 & 1\\
        O$_1$  & $4c$ & 0.09928(9) & $1/4$ & 0.74208(17) & 0.61(3) & 0.96(4) & 0.40(2) & 0 & 0.00(2) & 0 & 0.999(9)\\
        O$_2$  & $4c$ & 0.45275(9) & $1/4$ & 0.20116(18) & 0.28(3) & 0.92(4) & 0.68(3) & 0 & 0.00(2) & 0 & 1.016(8)\\
        O$_3$  & $8d$ & 0.16571(6) & 0.04314(12) & 0.27867(13)  & 0.66(3) & 0.75(3) & 0.58(2) & 0.17(2) & 0.10(2) & 0.03(2) & 1.008(6)\\
        \hline
    \end{tabular}
\end{table*}

The temperature dependency of the neutron diffraction intensity of 8 different magnetic Bragg peaks is shown in Fig. \ref{fig:E5complete}. Some peaks, such as $(0,1,0)$, display a power law behavior with onset at around $21\,\mathrm{K}$. Other peaks, like $(0,0,-1)$, display a linear temperature dependence with an onset temperature close to $25\,\mathrm{K}$. Finally, peaks such as $(3,0,-1)$, display both trends. All peak intensities were fitted to the following function:
\begin{widetext}
\begin{align}
	I(T) = \left\lbrace
	\begin{matrix}
		B					    & \quad \mathrm{for} & \quad T_2 < T\\
		I_2 \ (T_2 - T) + B	& \quad \mathrm{for} & \quad T_1 < T \leq T_2\\
		I_1 \ (T_1 - T)^{2\beta} + I_2 \ (T_2 - T) + B	& \quad \mathrm{for} &\quad T \leq T_1,\\
	\end{matrix} \right.
	\label{eq:e5fit}
\end{align}
\end{widetext}
where $T_2 > T_1$ are the transitions temperatures, $\beta$ is the critical exponent for the power law behavior and $I_1$, $I_2$ and $B$ are constants. $T_1$ and $T_2$ are fitted globally whereas $\beta$, $I_1$, $I_2$ and $B$ vary between data set. The obtained transition temperatures are $T_2 = 25.7(2)\,\mathrm{K}$ and $T_1 = 20.8(1)\,\mathrm{K}$ with the average value of $\beta = 0.32(3)$. Our data is generally well described with this model, see Fig. \ref{fig:E5complete}.

From the 8 magnetic Bragg peaks we can refine the magnetic structure at base temperature and obtain a combination of a $C$-type structure with moment components along $a$ and $b$ as well as a smaller $A$-type component with moments along $c$ with $R_F = 0.132$. The ordered moment is $\mu = (2.23(2),1.61(6),0.33(3))\mu_B$ which yields an angle in the $(a,b)$-plane of $\varphi = 54(1)^{\circ}$. This is consistent with the angle of $\varphi \approx 60^{\circ}$ as obtained from the subsequent polarized neutron scattering experiment. In the polarized experiment we observed a possible change in crystal symmetry at $T_1$ and the results were inconclusive with respect to the existence of a spin component along $c$. At E5, we obtained a sizeable component along $c$ with $A$-type symmetry like in the parent compound, \Ni\ \cite{toftpetersen2011}. Our observations are therefore contradictory and inconclusive on this point.

The total ordered moment is $|\mu| = 2.76(3)\,\mathrm{\mu_B}$. For \Fe\ and \Ni\ the measured ordered moments are respectively $4.2\,\mathrm{\mu_B}$ \cite{rousse2003} and $2.2\,\mathrm{\mu_B}$ \cite{jensen2009_struc} per magnetic ion yielding an expected moment of $2.6\,\mathrm{\mu_B}$ for the chemical composition of the sample. That is in very good agreement with the refinement result.

For the phase at intermediate temperatures, $T_1 \leq T \leq T_2$, it was not possible to unambigiously determine the magnetic structure from our E5 data. However, the results indicate a $C$-type structure with moments along $b$ and $|\mu| = 1.27\,\mathrm{\mu_B}$.

\begin{figure}[t!]
	\centering
	\includegraphics[width = 0.9\columnwidth]{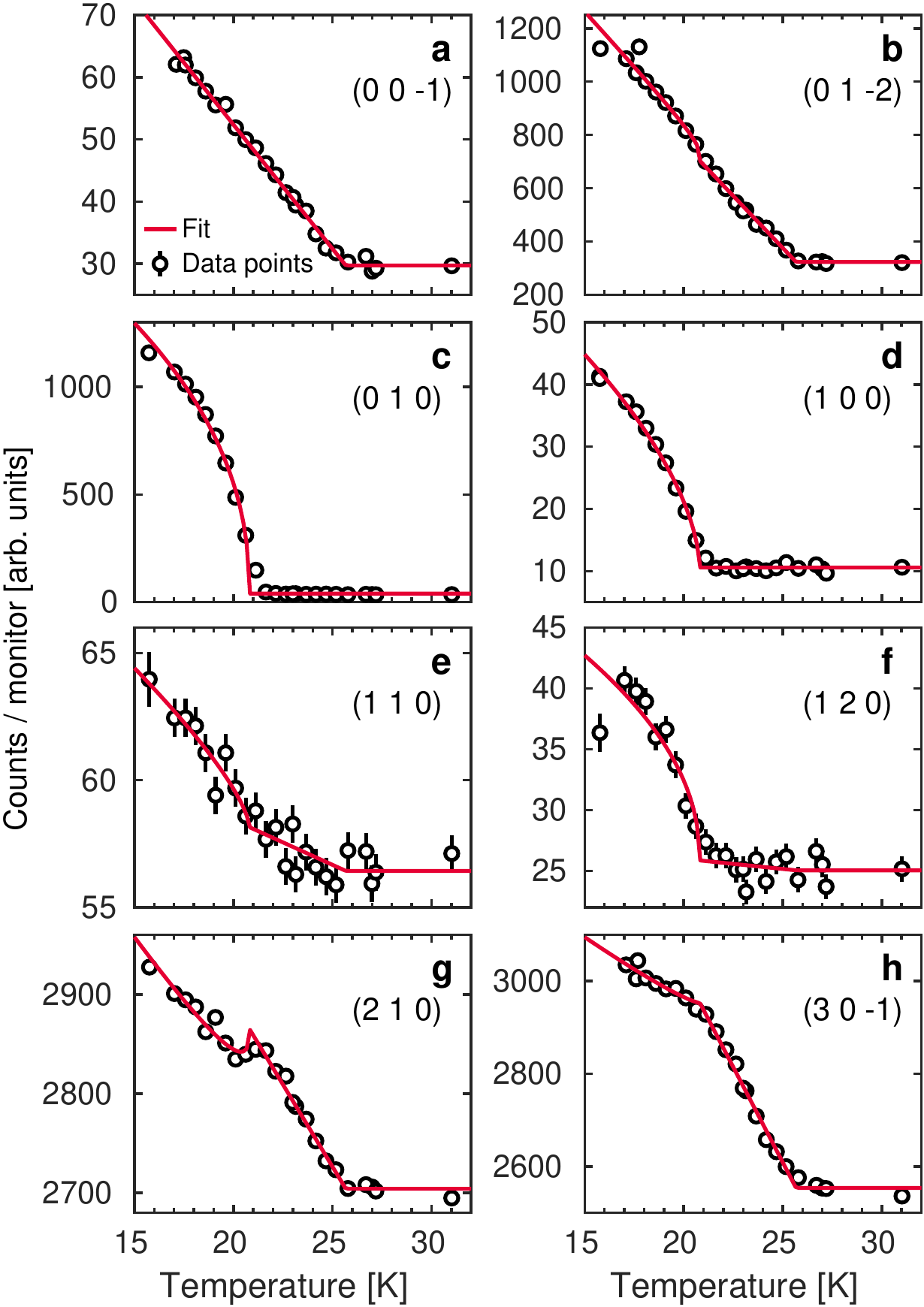}
	\caption{\textbf{Neutron intensity as a function of temperature collected at E5.} Data points are shown with open circles and the solid red lines are the global fit to Eq. \eqref{eq:e5fit}. Two transitions are clearly observed.}
	\label{fig:E5complete}
\end{figure}

\begin{figure*}[t!]
	\centering
	\includegraphics[width = 0.8\textwidth]{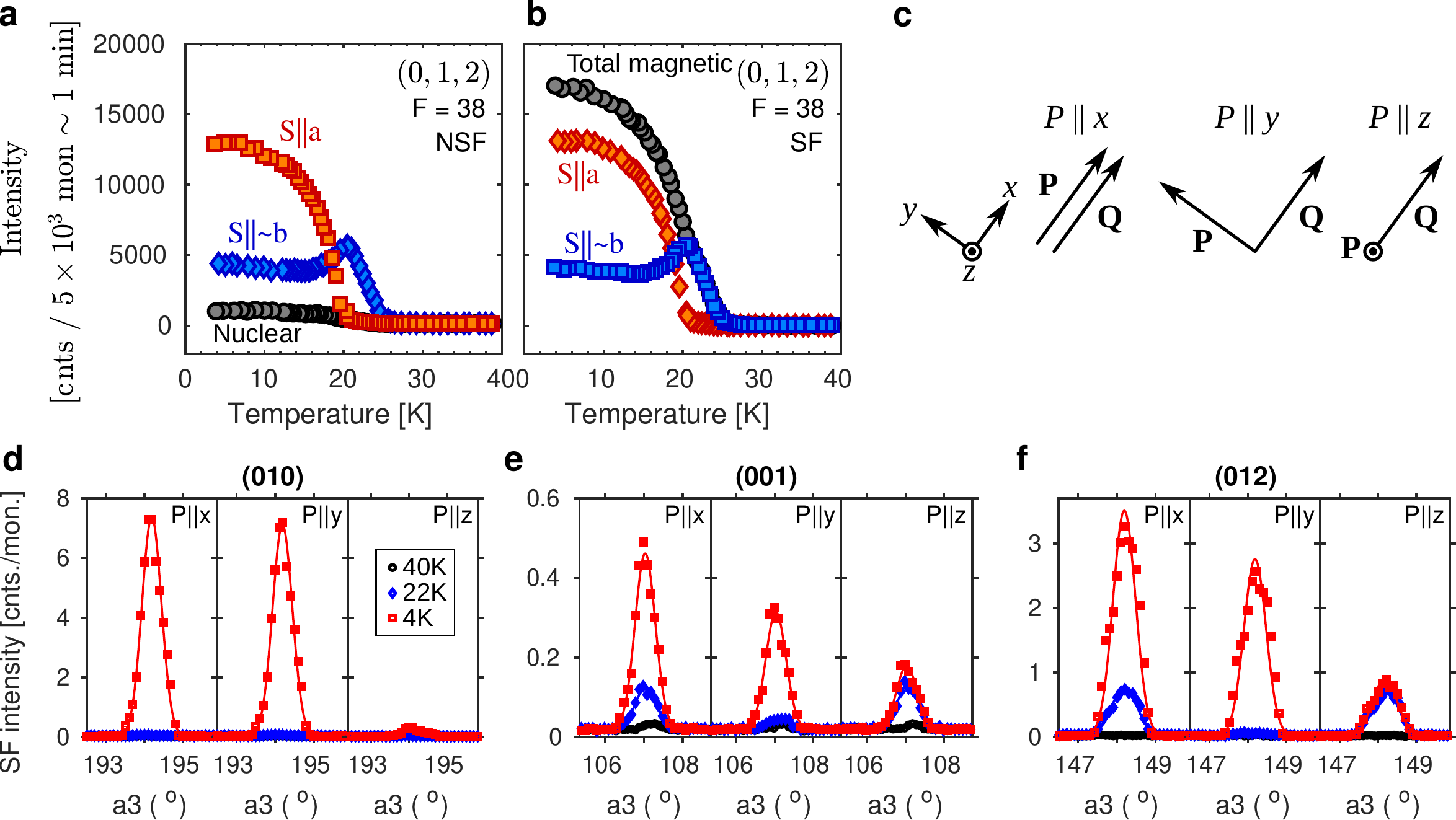}
	\caption{\textbf{Polarized neutron diffraction data collected at 4F1.} The intensities as a function of temperature for $(0,1,2)$ in panels \textbf{a}-\textbf{b} clearly show two transitions. \textbf{c}, The coordinate system used in the analysis of the polarized neutron diffraction data \cite{moon1969}. Panels \textbf{d}-\textbf{f} show rocking curves of the Bragg peaks respectively $(0,1,0)$, $(0,0,1)$ and $(0,1,2)$ at selected temperatures. Solid lines are Gaussian fits to the data.}
	\label{fig:4F1scans}
\end{figure*}

\medskip
\noindent {\bf 4F1 experiment.} In addition to data for the $(0,1,0)$ and $(0,0,1)$ magnetic Bragg peaks presented in the main text we also measured $(0,1,2)$ using polarized neutron diffraction at 4F1, see Figs. \ref{fig:4F1scans}a-b. The SF intensity as collected in rocking scans for the three peaks are shown in panels d-f at selected temperatures. The Bragg peaks are resolution limited signifying long-range magnetic order in the system. The intensity ratio and overall resemblance of the data for $(0,0,1)$ and $(0,1,2)$ consolidate that the main magnetic structure component is $C$.


The angle, $\varphi$ in the $(a,b)$-plane is calculated as:
\[
	\tan \varphi = \sqrt{\frac{I_a}{I_b}},
\]
where $I_a = I_y^{\mathrm{NSF}} -  I_x^{\mathrm{NSF}}$ or $I_a = I_x^{\mathrm{SF}} -  I_y^{\mathrm{SF}}$ and $I_b = I_z^{\mathrm{NSF}} -  I_x^{\mathrm{NSF}}$ or $I_b = I_x^{\mathrm{SF}} -  I_z^{\mathrm{SF}}$. 
$I_b$ for $(0,1,2)$ is corrected for the rotation between the sample coordinate system and the crystallographic axes, see Fig. \ref{fig:4F1scans}c. Form factors and Lorentz factors may be omitted when regarding Bragg peaks individually like here. At the lowest probed temperatures we obtain canting angles respectively for $(0,1,2)$ and $(0,0,1)$ of $\varphi = 60.9(5)^{\circ}$ and $56.1(9)^{\circ}$.

\bigskip
\noindent
{\bf \large Pyrocurrent measurements}
\medskip

\begin{figure}
	\centering
	\includegraphics[width = 0.94\columnwidth]{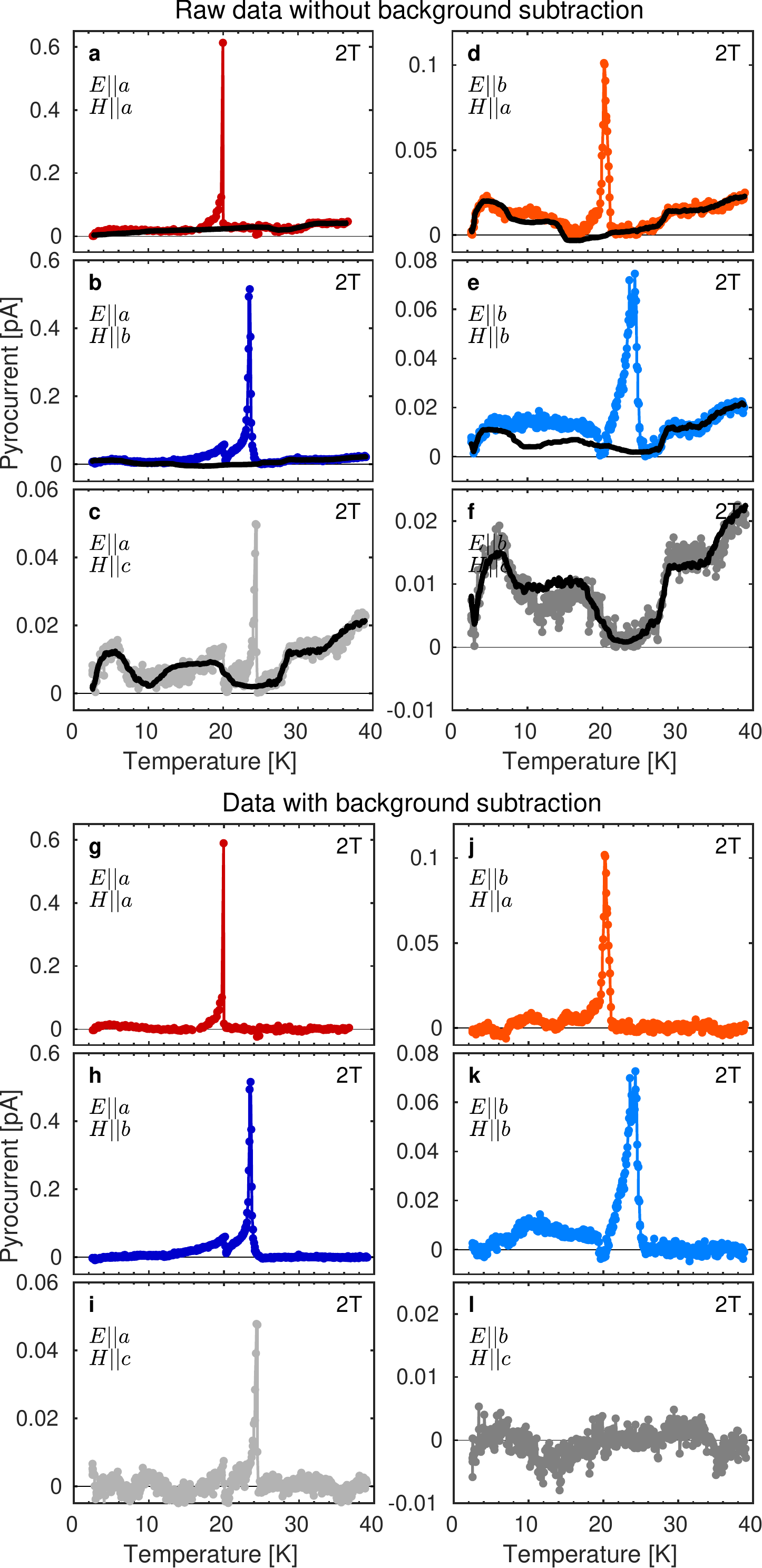}
	\caption{\textbf{Background subtraction for the pyrocurrent measurements.} Top panels show the raw data (colored dots) with the background plotted on top (black curves). Bottom panels show the data after background subtraction.}
	\label{fig:pyrocurrentBackground}
\end{figure}

\noindent The pyrocurrent in LiNiPO$_4$, LiFePO$_4$ and LiNi$_{0.8}$Fe$_{0.2}$PO$_4$ was measured using the quasi-static method \cite{chynoweth1956}. 
The electric polarization, $P(T_1)$ at temperature $T_1$, was calculated from the measured pyrocurrent, $I$, using following formula:
\[
	P (T_1) = P(T_0) - \frac{1}{A} \int_{T_0}^{T_1} I(T) \left( \frac{dT}{dt} \right)^{-1} dT,
\]
where $\frac{dT}{dt}$ is the temperature ramp rate and the minus sign in front of the integral is chosen such that $\Delta P>0$ upon heating, as in our case. The constant, $P(T_0)$, is chosen such that the polarization is zero in the paramagnetic phase. The integration is performed numerically. 
Since the polarization is calculated from the current, a small constant leak current in the measurement system will result in an overall slope in the polarization. This is relatively simple to correct for but any temperature dependent background signal may introduce features in the polarization that do not originate from the sample. In order to describe the background as well as possible, we construct a dataset based on the zero-field data and parts of the in-field data where it is known that there should be no pyrocurrent signal. At zero field, the lithium orthophosphates do not support the ME effect. The background is different for each electric and magnetic field direction. The pyrocurrent signal before and after background subtraction is shown in Fig. \ref{fig:pyrocurrentBackground}. It is clear that above the transition temperature we obtain a flat pyrocurrent signal as desired.

Since the measured pyrocurrent is the derivative of the polarization, this particular experimental method is better suited for characterizing second order phase transitions. First order phase transitions are identified by the step-like onset in the order parameter, i.e. in our case the electric polarization. Consequently the derivative at the transition is described by a delta function and is experimentally challenging to capture. Therefore, the pyrocurrent method is not a good choice for characterizing first order transitions such as in LiNiPO$_4$ \cite{vaknin2004}. On the other hand, it works well for second order transitions where the derivative of the order parameter is well-defined at all temperatures as is the case for LiFePO$_4$ \cite{toftpetersen2015} and LiNi$_{0.8}$Fe$_{0.2}$PO$_4$.

\bigskip
\noindent
{\bf \large Monte Carlo simulations} \newline
In the main part of the manuscript we use a simplified set of parameters for our Monte Carlo simulations but for completeness the full set of established magnetic exchange couplings and single ion anisotropies for both parent compounds, LiNiPO$_4$ and LiFePO$_4$, are given in Table \ref{tab:parameters}. We performed additional simulations using the full parameter set but found qualitatively similar behavior. For those simulations we used the mean to describe inter-species couplings: $J^{\mathrm{Ni/Fe}} = \frac{1}{2} \left( J^{\mathrm{Ni}} +  J^{\mathrm{Fe}} \right)$.

\begin{table}[b!]
	\caption{Exchange and single-ion anisotropy constants for \Ni\ \cite{jensen2009_dyna} and \Fe\ \cite{toftpetersen2015} given in milli electronvolt.}
	\label{tab:parameters}
	\small
	\begin{tabular}{c c c c c c c c}
		\\
		\Ni \\
		\hline
		$J_{bc}$	& $J_b$	& $J_c$	& $J_{ab}$	& $J_{ac}$	& $D^a$	& $D^b$\\
		1.04(6)		& 0.670(9)	& -0.05(6)	& 0.30(6)		& -0.11(3)		& 0.339(2)	& 1.82(3)\\ \\
		\Fe \\
		\hline
		$J_{bc}$	& $J_b$	& $J_c$	& $J_{ab}$	& $J_{ac}$	& $D^a$	& $D^c$\\
		0.77(7)		& 0.30(6)	& 0.14(4)	& 0.14(2)		& 0.05(2)		& 0.62(12)	& 1.56(3)
	\end{tabular}
\end{table}

\medskip
\noindent {\bf Susceptibility for $x=0.06$.}
The main text of this paper is focused on intermediate ranges of $x$ where \NiFex\ hosts an oblique phase for $0.1 < x < 0.6$. The measured susceptibility for $x=0.20$ is well reproduced by our Monte Carlo simulations, as seen by comparing Figs. \ref{fig:phasediagram_susceptibility}c and \ref{fig:simulation}a in the main text. To further illustrate the success of our Monte Carlo simulations we compare the the measured and simulated susceptibility for LiNi$_{0.94}$Fe$_{0.06}$PO$_4$ in Fig. \ref{fig:chiLowFe}. The experiment was performed using a Cryogenic Ltd. cryogen free vibrating sample magnetometer and three separate crystals of masses $50$, $40$ and $20\,\mathrm{mg}$. A magnetic field of $0.5\,\mathrm{T}$ was applied along the respective crystallographic axes. The simulated susceptibility was calculated from simulations performed without a Zeeman term in the Hamiltonian, as described in the Methods section.

The experimental susceptibility curves for LiNi$_{0.94}$Fe$_{0.06}$PO$_4$ with $H||a$ and $H||c$ are similar to those for LiNiPO$_4$ (compare Fig. \ref{fig:chiLowFe}a with Fig. \ref{fig:phasediagram_susceptibility}d) indicating that the magnetic moments order along $c$ as in the parent compound. For $H||b$, however, the susceptibility of LiNi$_{0.94}$Fe$_{0.06}$PO$_4$ displays a clear upturn below $T_N \approx 17.4\,\mathrm{K}$ resembling a Curie tail associated with small amounts of paramagnetic impurities. This upturn is not present in \Ni. 

The simulated susceptibility curves shown in Fig. \ref{fig:chiLowFe}b are in remarkably good agreement with the experimental data including the low-temperature increase for $H||b$. The easy axis for LiFePO$_4$ is $b$ and locally this must also hold for the individual Fe$^{2+}$ ions in LiNi$_{0.94}$Fe$_{0.06}$PO$_4$. We speculate that a small fraction of such ions act as magnetic impurities that are effectively decoupled from the surrounding antiferromagnetically ordered Ni$^{2+}$ ions, and therefore behave as free magnetic ions upon applying a magnetic field along $b$.

Note that we do not observe the incommensurate phase reported in LiNiPO$_4$ \cite{toftpetersen2011} for small $x$ in our simulations. However, our preliminary neutron diffraction study on LiNi$_{0.94}$Fe$_{0.06}$PO$_4$ shows that this phase persists and should be subject for future studies.

\noindent {\bf Local moment fluctuations.} Both our experiments and simulations show that there is magnetic long range order in the \NiFex\ system for all $x$. However, in the simulations we also see local fluctuations of the moment direction as illustrated in Fig. \ref{fig:MCstructures}a. The site-dependent single-ion anisotropy means that the average direction of the moments vary depending on whether there is Ni or Fe on a specific site and which ion species reside on the neighboring sites, i.e. the local crystal field. In Fig. \ref{fig:MCstructures}b-c we plot the spin components along $a$, $b$ and $c$ for the Ni and Fe sites individually. At small $x$, the Ni spins are ordered along $c$ and for large $x$ along $b$ but with a crossover region with a large $a$-axis component. The picture is similar for the Fe spins except for small $x$ where the spin component along $b$ increases upon decreasing the Fe content. This is a result of the strong single-ion anisotropy of Fe and its large spin. Note that none of the spin components are zero due to the finite temperature in the simulations. A diffuse neutron scattering experiment would illuminate the local variation in moment orientation.

\begin{figure}
	\centering
	\includegraphics[width = 0.8\columnwidth]{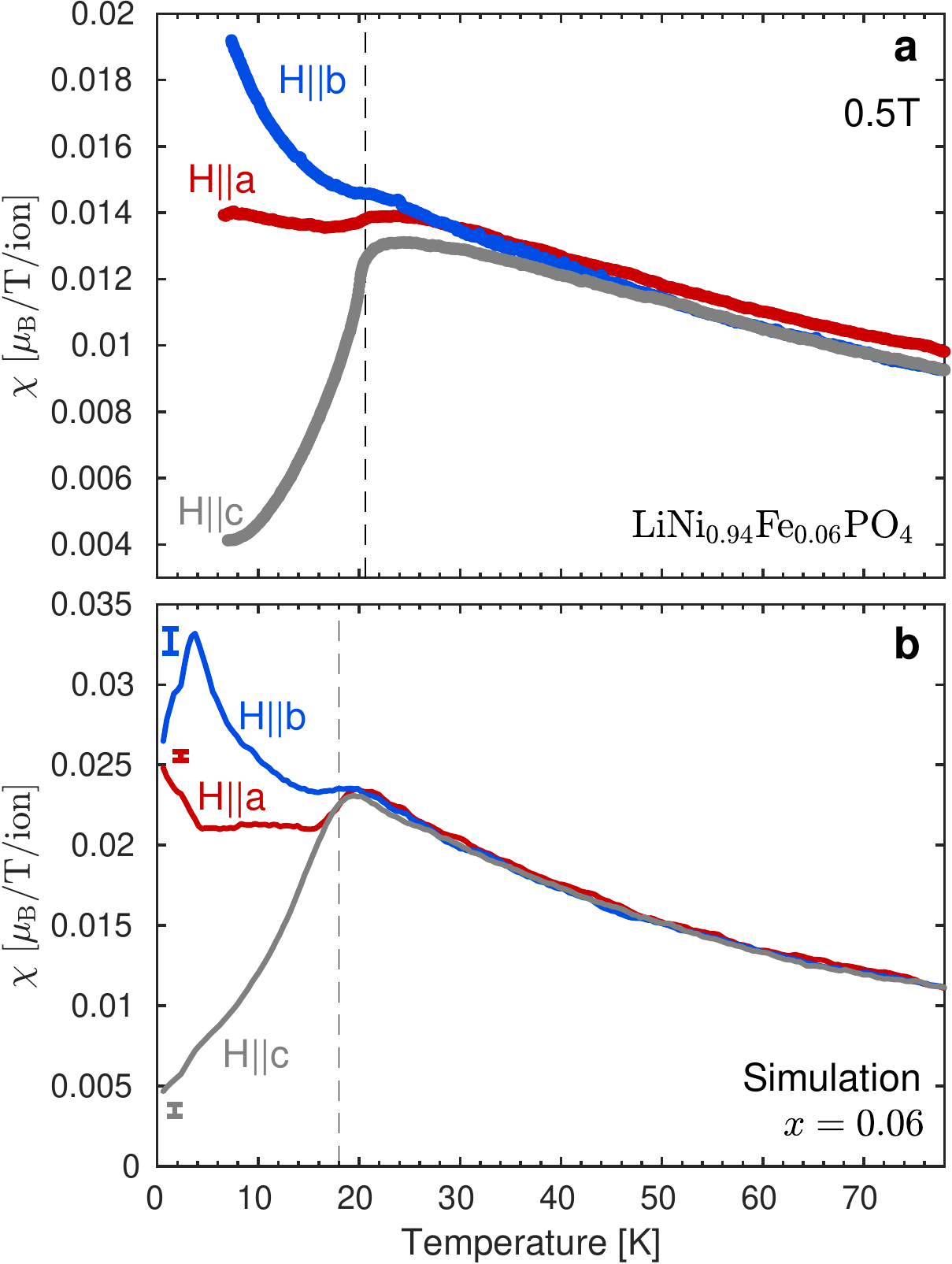}
	\caption{\textbf{Susceptibility for LiNi$_{0.94}$Fe$_{0.06}$PO$_4$,} Magnetic susceptibility measured in an applied field of $0.5\,\mathrm{T}$ (\textbf{a}) and simulated for $x = 0.06$ (\textbf{b}). Transition temperatures are marked by vertical dashed lines. The largest errors estimated for the simulations are shown on the left in panel \textbf{b}.}
	\label{fig:chiLowFe}
\end{figure}

\begin{figure*}
	\centering
	\includegraphics[width = 0.8\textwidth]{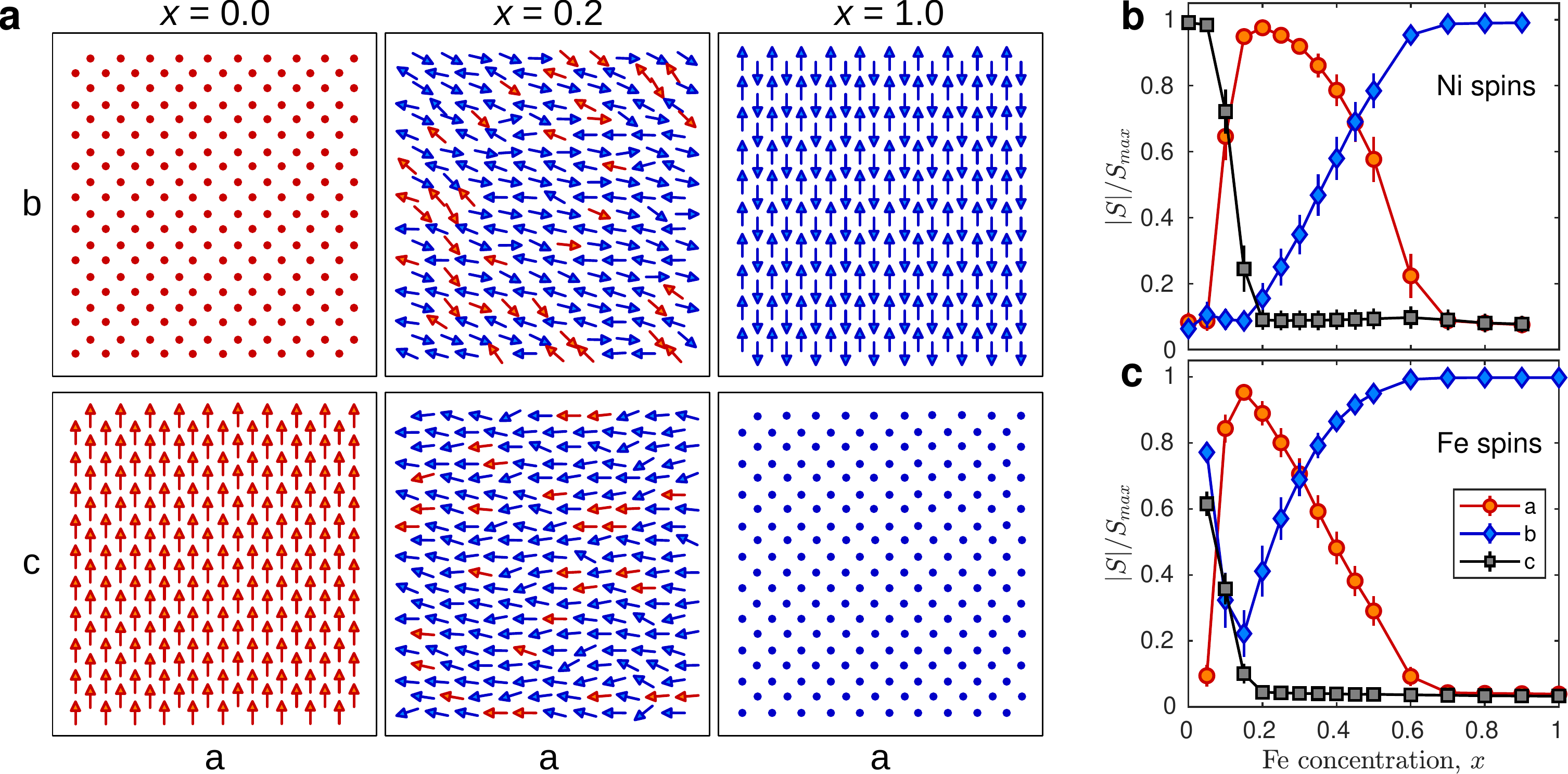}
	\caption{\textbf{Fluctuation of local moments.} \textbf{a}, Examples of system configurations for $x = 0.0, 0.2$ and $1.0$ shown in the $(a,b)$ and $(a,c)$ planes (top and bottom row respectively). All moments are shown with the same length to clearly show their orientation. Red arrows correspond to Ni$^{2+}$ and blue arrows to Fe$^{2+}$. In panels \textbf{b}-\textbf{c} are plots of the average of the absolute value of the spin components along the $a$, $b$ and $c$ directions for Ni and Fe sites.}
	\label{fig:MCstructures}
\end{figure*}

\end{document}